\documentclass[11pt]{article}
\usepackage[a4paper, portrait, margin=1.1811in]{geometry}

\usepackage[english]{babel}
\usepackage[utf8]{inputenc}
\usepackage[T1]{fontenc}
\usepackage{helvet}
\usepackage{etoolbox}
\usepackage{graphicx}
\usepackage{titlesec}
\usepackage{caption}
\usepackage{booktabs}
\usepackage{xcolor} 
\usepackage[colorlinks, citecolor=cyan]{hyperref}
\usepackage{caption}
\graphicspath{ {./images/} }
\usepackage{scrextend}
\usepackage{fancyhdr}
\usepackage{graphicx}
\usepackage{cite}

\usepackage{amsmath,amssymb,mathrsfs}
\usepackage{epsfig, graphicx}
\usepackage{epstopdf} 
\usepackage{url}
\usepackage{authblk}

\newtheorem{theorem}{Theorem}[section]

\newcommand{\be}{\begin{equation}}
	\newcommand{\ee}{\end{equation}}

\definecolor{ggreen}{cmyk}{1,  0,  1, 0}

\fancypagestyle{plain}{
	\fancyhf{}

}

\makeatletter
\patchcmd{\@maketitle}{\LARGE \@title}{\fontsize{16}{19.2}\selectfont\@title}{}{}
\makeatother

\newsavebox\affbox

\allowdisplaybreaks

\author[1]{\textbf{G. Martal\`o}}
\author[2]{\textbf{A.~J. Soares}}
\author[3,1]{\textbf{R. Travaglini}}
\affil[1] {Department of Mathematical, Physical and Computer Sciences, 
	University of Parma, Parco Area delle Scienze 53/A,  43124,
	Parma, Italy}
\affil[2]{Centre of Mathematics of the University of Minho, Campus de Gualtar,
	4710-057 Braga, Portugal}
\affil[3] { Istituto Nazionale di Alta Matematica ``Francesco Severi'', Piazzale Aldo Moro 5,  00185, Roma, Italy}

\titlespacing\section{0pt}{12pt plus 4pt minus 2pt}{0pt plus 2pt minus 2pt}
\titlespacing\subsection{12pt}{12pt plus 4pt minus 2pt}{0pt plus 2pt minus 2pt}
\titlespacing\subsubsection{12pt}{12pt plus 4pt minus 2pt}{0pt plus 2pt minus 2pt}

\titleformat{\section}{\normalfont\fontsize{10}{15}\bfseries}{\thesection.}{1em}{}
\titleformat{\subsection}{\normalfont\fontsize{10}{15}\bfseries}{\thesubsection.}{1em}{}
\titleformat{\subsubsection}{\normalfont\fontsize{10}{15}\bfseries}{\thesubsubsection.}{1em}{}

\titleformat{\author}{\normalfont\fontsize{10}{15}\bfseries}{\thesection}{1em}{}

\title{\textbf{A BGK-type model for multi-component gas mixtures undergoing a bimolecular chemical reaction}}

\date{}

\begin{document}

	\newpage
	\setcounter{page}{1}
	\renewcommand{\thepage}{\arabic{page}}
	
	\maketitle

	\noindent\rule{15cm}{0.5pt}
	
	
	\begin{abstract}
		\textbf{Abstract.}
		We {propose a new} kinetic BGK-type model for a mixture of four monatomic
		gases, undergoing a bimolecular and reversible chemical reaction. 
		The elastic and {reactive} interactions are described separately by {distinct} relaxation terms
		{and} the mechanical operator is the sum of binary BGK contributions, one for each pair of interacting species.
		In this way, our model separately incorporates the effects of mechanical processes and chemical reactions.
			Additionally, it retains the effects of inter-species interactions which are proper of the mixture.
		The dependence of Maxwellian attractors on the main macroscopic fields is explicitly expressed by assuming that the exchange rates for momentum and energy of mechanical and chemical operators coincide with the ones of the corresponding Boltzmann terms. Under suitable hypotheses, the relaxation of the distribution functions to equilibrium is shown through entropy dissipation. Some numerical simulations are included to investigate the trend to equilibrium.
		\\ \\
		\textbf{\textit{Keywords}}: \textit{Kinetic theory; BGK equations; Reactive mixtures; Exchange rates}
		\\ \\
		\textbf{\textit{AMS subject classification 2020}}: {82C40; 76P05; 80A30; 35Q20 }
	\end{abstract}
	
	\noindent\rule{15cm}{0.4pt}
	
	
	
	\section{Introduction}
	\label{sec:int}
	
	Kinetic theory provides a natural framework to describe the evolution of rarefied gas mixtures. 
	The classical description in the kinetic setting is given by Boltzmann equations that describe the evolution of the distribution functions. 
	This formulation, though, results in being very complicated to deal with, due to the presence of {integral collision operators with quadratic non-linearities}. 
	Indeed, {the Boltzmann equation} is well known in literature \cite{Cercignani,Chapman-Cowling,Giovangigli}, but the investigation of its mathematical properties \cite{Briant-Daus,Gamba-Pavic}, as well as the construction of efficient numerical schemes \cite{Bondesan-Boudin-Grec,Crestetto-Klingenberg-Pirner,Wu,Zhang}, {continues to motivate challenging problems that are} still in progress. 
	
	Some alternative approaches have been proposed to overcome {the} difficulties {associated with the integral collision term}.
	The most famous one is based on the pioneering paper {by} Bhatnagar, Gross, and Krook \cite{BGK54}. 
	In this work, the authors suggested replacing the Boltzmann collision operator with a linear non-integral term, prescribing the relaxation towards 
	an equilibrium Maxwellian attractor. 
	Being the description formulated for a single gas, consistent generalizations to gas mixtures were highly desirable, 
	especially concerning their application to physical processes.  
	Here, by consistent generalizations we mean models that ensure the correct conservation laws, satisfy an $\cal H$-theorem, and are compatible with the proper characterization of the equilibrium states.
	
	The first consistent model has been built up in \cite{AAP}, where the authors assume that, in the evolution equation of each component, 
	the collision phenomenon is governed by a unique relaxation operator, which collects the effects of the interactions 
	among all the constituents.
	
	It is essential to point out that the extension {of the BGK model} to a gas mixture is not unique. 
	In the BGK attractors, indeed, several free parameters intervene, {and they} can be suitably chosen 
	to reproduce some basic properties of the original Boltzmann operators. 
	For this reason, in the same spirit, different formulations can be found in the literature for inert and reactive mixtures of monatomic and polyatomic gases 
	\cite{BGS-PhysRevE,Bisi-Monaco-Soares,BT1,BT2, BT3,Brull-Schneider-CMS2014,garzo1989kinetic,Haack-Hauck-Murillo,Klingenberg-Pirner-Puppo}.
	
	Among the others, we refer to \cite{BBGSP}, where the proposed model of BGK-type {for inert mixtures} mimics the same structure of the Boltzmann one, 
	i.e. the collision operator is the sum of binary terms, each one accounting for the exchanges, occurring via mechanical interactions, of any pair of components. This formulation has the great advantage of distinguishing the different collision types and, consequently, 
	allows the investigation of some multiple-scale {regimes \cite{bisi2021macroscopic,Giovangigli}}, 
	which are often present in plasma physics. For example, as described in \cite{Galkin}, in the presence of heavy and light components in the mixture, momentum and energy exchanges proceed at different scales, according to the mass ratio.
	A typical example is given by a mixture of ions and electrons, as detailed in \cite{Pirner}. 
	From this perspective, recently a class of mixed Boltzmann-BGK models has been presented \cite{Ferrara,BGLM}, 
	where the accuracy of the Boltzmann operator is required to describe the dominant phenomenon, 
	while the slow process is modeled by more manageable BGK terms.
	
	Another suitable extension concerns the possibility to include nonconservative interactions, like chemical reactions. In fact, the capability of modeling gas mixtures in a reactive framework has a significant value for real-world applications, like re-entry problems in the atmosphere \cite{Li-Zhang,Baranger-etal} and industrial processes \cite{Cercignani-Frezzotti-Lorenzani,Lorenzani-2019}.
		
		While kinetic equations were initially extended to inert gas mixtures, allowing for elastic binary interactions between different components \cite{Cercignani,Chapman-Cowling}, there has long been a growing interest in chemical phenomena, in order to provide more realistic models in gas dynamics. Since the 90s, several models based on Boltzmann formulation have included reversible chemical reactions, with four monatomic gases undergoing a bimolecular reversible reaction.
		Such models can be found in \cite{rossani1999note}, as by-product
		of previous works by the authors.
		
		The extension of BGK models to chemically reacting mixtures is a very challenging topic. Significant efforts have been made in the last years, for example, in \cite{BGS-PhysRevE,GS2004}, a unique relaxation operator can take account of both mechanical and chemical interactions. Recently, some tentatives of extension to reacting mixtures of monatomic and polyatomic gases have been presented \cite{BT2,BT3, Bisi-Monaco-Soares}.
		
		On another research line, other formulations have been proposed for reactive mixtures, 
		separating the chemical contribution from the mechanical one \cite{BPS2014,Kremer-PandolfiBianchi-Soares-PoF2006}. 
		In paper \cite{BPS2014}, the authors collect all mechanical contributions
			in one single BGK operator, whereas
			in paper \cite{Kremer-PandolfiBianchi-Soares-PoF2006}, one individual BGK operator
			is introduced for each Boltzmann operator.
			Both approaches propose innovative models. However, some limitations can be identified. 
			The model developed in \cite{BPS2014} loses the mixture effects to some extent,   
			and fails to account for common processes among the constituents, 
			such as exchanges of momentum and energy, and cross-diffusion, for example.
			The model developed in \cite{Kremer-PandolfiBianchi-Soares-PoF2006} captures
			all the mixture effects but does not satisfy an $\cal H$-theorem, representing 
			a significant weakness from the mathematical point of view.
		
		In this manuscript, we 
		follow the same line and aim to address this gap,
			by developing a model that satisfies an $\cal H$-theorem and ensures the correct characterization of
			mechanical and chemical equilibrium states.	
			Specifically, we intend to extend the mechanical BGK model proposed in \cite{BBGSP},
		with one BGK operator for each Boltzmann term,
		by adding a BGK reactive operator describing the chemical effects.
		
		Thus, we consider a mixture of four monatomic gases undergoing binary elastic collisions and a bimolecular reversible reaction, 
		involving all the components. Our approach separates the mechanical and the chemical operators. 
		The former is the sum of binary terms, as given in \cite{BBGSP},
		whereas the latter is of BGK type, whose auxiliary parameters have to be determined properly. 
		The chemical term differs from the ones in \cite{BPS2014}, where all auxiliary Mawellians share the same 
		mean velocity and temperature, and from those proposed in \cite{Kremer-PandolfiBianchi-Soares-PoF2006}, 
		where authors assume a proper perturbation of Maxwellian attractors. 
		In our case, each auxiliary Maxwellian has its own fictitious mean velocity and temperature, depending on the species features.
		
		This formulation allows for investigating the behavior of each component, 
		as well as the trend to equilibrium of its observable quantities, 
		which can differ from species to species because of their distinct masses. 
		Moreover, the model captures the different time scales at which mechanical and chemical phenomena proceed,
		being crucial to describe different evolution regimes for what concerns
			mechanical processes and chemical reactions.

	As done for the inert {mixture}, the collision operator for each involved species is given by a sum of BGK terms, 
	each {term being} distinguished by its specific set of parameters. Assuming that the exchange rates for momentum and energy of BGK mechanical and chemical operators coincide with the corresponding exchange rates of each Boltzmann integral operator, 
	we can ensure the { consistency} of the model.
	{The extension of the model to a reactive mixture represents a non-trivial problem 
		because the rearrangement of masses and redistribution of energy among the reactive species
		lead to cumbersome computations.}
	
	{Assuming} the Maxwell molecule intermolecular potential, it is possible to explicitly calculate the production terms in the mechanical contributions for both Boltzmann and BGK formulations. This results in an exact relationship between the parameters of local attractors and the macroscopic fields of each species. Regarding chemical contributions, a reasonable input distribution function is required to compute the production terms in the Boltzmann setting. 
	To this aim, we approximate the  distribution function  by a proper perturbation of the Maxwellian function {in the constituent reference frame}, using an expansion analogous to the one introduced in \cite{Kremer-PandolfiBianchi-Soares-PoF2006}, where it is used to model the auxiliary distribution in the BGK setting instead.
	In other words, it results in a linearization with respect to species velocities and temperatures and corresponds to a 
	not-so-far-away deviation from the mechanical equilibrium. Other choices are admissible, like the Grad-type approximation leading to 13-moment equations in the hydrodynamic limit \cite{BisiCMT2002}.
	

	After this introduction, the paper is organized as follows.
	{In Section \ref{sec:descmodel}, a}fter recalling the classical Boltzmann equation for a reactive mixture of monatomic gases, 
	we {propose a new} BGK-type model, by introducing a kinetic operator as the sum of relaxation terms. 
	In Section \ref{sec:par}, we explicitly compute the auxiliary parameters in the Maxwellian attractors of BGK terms. 
	This can be done by assuming that the exchange rates for Boltzmann and BGK operators are the same for any type of interaction. 
	The relaxation of the distribution functions towards the mechanical and chemical equilibrium, is studied in Section \ref{sec:Htheo}. In Subsection \ref{ssec:Htheo}, we show the existence of an $H$-functional providing the entropy dissipation, under the hypotheses of species velocity equalization and the mass action law. In Subsection \ref{ssec:numerics}, with reference to a real chemical reaction scenario, some numerical simulations are performed to show  the asymptotic approach to the equilibrium.
	Lastly, some concluding remarks are given in Section {\ref{sec:conc}}.
	
	
	\section{Kinetic equations}
	\label{sec:descmodel}
	
	We consider a reactive mixture of four monatomic rarefied gases $G_i$, $i=1,2,3,4$, 
	undergoing elastic mechanical collisions and a bimolecular reversible chemical reaction {of type}
	\begin{equation}
		G_1+G_2\leftrightarrows G_3+G_4\,.
		\label{reaction}
	\end{equation}
	Each $i-$ component of the mixture is endowed with its own {molecular} mass $m_i$ and internal energy $E_i$, 
	such that $\Delta E = E_3 + E_4 - E_1 - E_2 > 0$, meaning that the forward reaction is endothermic.
	
	Let $f_i=f_i(\mathbf{x},\mathbf{v},t)$ be the distribution function of the particles of the $i-$th component in the phase space, 
	depending on space variable $\mathbf{x}\in\mathbb{R}^3$, microscopic velocity $\mathbf{v}\in\mathbb{R}^3$ 
	and time $t\in\mathbb{R}_+$.
	We use the notation $\mathbf{f}=[f_1,f_2,f_3,f_4]$ for the vector collecting all the distribution functions.
	The time-space evolution of each function is governed by a proper kinetic equation
	\begin{equation}
		\dfrac{\partial f_i}{\partial t}+\mathbf{v}\cdot\nabla_{\mathbf{x}}f_i=\mathcal{Q}_i[\mathbf{f}]\,,\quad i=1,2,3,4\,,
		\label{kin_eq}
	\end{equation}
	where the operator $\mathcal{Q}_i$, at the right-hand side, takes into account both mechanical and chemical interactions.
	
	\medskip
	
	For each component of the mixture, we define the main macroscopic fields as suitable moments of the distribution function. 
	In particular, the number density $n_i$, mean velocity $\mathbf{u}_i$ and temperature $T_i$ 
	for the $i-$th component are computed as
	\begin{equation}
		n_i = \int_{\mathbb{R}^3}f_i(\mathbf{v})d\mathbf{v} \,, \quad
		\mathbf{u}_i = \dfrac{1}{n_i}\int_{\mathbb{R}^3}\mathbf{v}f_i(\mathbf{v})d\mathbf{v} \,, \quad 
		T_i = \dfrac{m_i}{3n_i}\int_{\mathbb{R}^3}|\mathbf{v}-\mathbf{u}_i|^2f_i(\mathbf{v})d\mathbf{v}\,.
	\end{equation}
	Mixture observable quantities are {defined} as a combination of the corresponding species ones, as
	\begin{equation}
		\begin{aligned}
			n = \sum_{i=1}^4 & n_i\,,\quad
			\rho=\sum_{i=1}^4\rho_i=\sum_{i=1}^4m_in_i\,,\quad
			\mathbf{u} = \dfrac{1}{\rho}\sum_{i=1}^4\rho_i\mathbf{u}_i \,,\\
			T&=\dfrac{1}{n}\left[\sum_{i=1}^4n_iT_i + \dfrac13\sum_{i=1}^4m_in_i\left(u_i^2-u^2\right)\right] .
		\end{aligned}
	\end{equation}
	
	
	\subsection{Boltzmann formulation}
	\label{ssec:BE}
	
	The 
	{more relevant equation in Kinetic Theory} is the one based on the Boltzmann formulation.
	The collision operator is split into a sum of two contributions
	\begin{equation}
		\mathcal{Q}_i^{Bol} = \widetilde{\mathcal{Q}}_i^{Bol} + \widehat{\mathcal{Q}}_i^{Bol}\,,
		\label{eq:QB}
	\end{equation}
	where $\widetilde{\mathcal{Q}}_i^{Bol}$ takes {into} account {the} mechanical encounters, 
	{whereas} $\widehat{\mathcal{Q}}_i^{Bol}$ considers the effects of the chemical reaction. 
	As usual in Boltzmann formulat{ion} for mixtures \cite{Cercignani,Chapman-Cowling,Giovangigli}, the mechanical contribution 
	is decomposed in {a} sum of binary terms, {each} one {describing the} 
	elastic interacti{ons between $i$ and $j$ components, namely} 
	\begin{equation}
		\widetilde{\mathcal{Q}}_i^{Bol}=\sum_{j=1}^4\widetilde{\mathcal{Q}}_{ij}^{Bol}\,.
		\label{Bolt_chem}
	\end{equation}
	{Both the} mechanical and chemical {terms are non-linear} operators of integral type.
	More precisely, the mechanical {terms} are given {by}
	\begin{equation}
		\widetilde{\mathcal{Q}}_{ij}^{Bol}=\iint_{\mathbb{R}^3\times\mathbb{S}^2}g\widetilde{\sigma}_{ij}(g,\widehat{\Omega}\cdot\widehat{\Omega}^\prime)\left[f_i(\mathbf{v}^\prime)f_j(\mathbf{w}^\prime)-f_i(\mathbf{v})f_j(\mathbf{w})\right]d\mathbf{w}d\widehat{\Omega}^\prime\,,\quad i,j=1,2,3,4\,,
		\label{Bolt_mech}
	\end{equation}
	where $\mathbf{g}=\mathbf{v}-\mathbf{w}=g\widehat{\Omega}$ denotes the relative velocity {of the colliding pair,
		with} $g$ {being} its modulus and $\widehat{\Omega}$ {the corresponding unit vector.
		Pre-} and post-collision velocities are related by the {conservation laws} of mass, momentum, and total energy 
	{during} the collision{s, that is}
	\begin{equation}
		\begin{aligned}
			m_i+m_j&=m_i+m_j\\
			m_i\mathbf{v}+m_j\mathbf{w}&=m_i\mathbf{v}^\prime+m_j\mathbf{w}^\prime\\
			\dfrac12m_iv^2+E_i\dfrac12m_jw^2+E_j&=\dfrac12m_i(v^\prime)^2+E_i+\dfrac12m_j(w^\prime)^2+E_j\,.
		\end{aligned}
	\end{equation}
	The quantity {$\widetilde{\sigma}_{ij}(g,\widehat{\Omega}\cdot\widehat{\Omega}^\prime)$ in \eqref{Bolt_mech}} 
	denotes the {elastic} cross-section \cite{Cercignani}, {and} takes {into} account the inter-molecular potential,
	and it is assumed to satisfy proper symmetry conditions
	\begin{equation}\label{symmetrycross}
		\widetilde{\sigma}_{ij}(g,\widehat{\Omega}\cdot\widehat{\Omega}^\prime)=\widetilde{\sigma}_{ij}(g,-\widehat{\Omega}\cdot\widehat{\Omega}^\prime)=\widetilde{\sigma}_{ji}(g,\widehat{\Omega}\cdot\widehat{\Omega}^\prime)\,.
	\end{equation}
	{In} this paper, we assume that {the mixture} particles interact according to {the} Maxwell molecule model \cite{maxwell1866xiii}, and {therefore}
	\begin{equation}
		\widetilde{\lambda}_{ij}^{(0)}
		= \int_{\mathbb{S}^2} g\widetilde{\sigma}_{ij}(g,\widehat{\Omega}\cdot\widehat{\Omega}^\prime)d\widehat{\Omega}^\prime
		= \text{constant}\,.
	\end{equation}
	
	{Concerning chemical interactions, they}, can {be} described in an analogous way, 
	by means of {a} Boltzmann-like operator. 
	{In} particular, the chemical contribution {to} the {kinetic equation of the $i-$th component} is given by \cite{rossani1999note}
	\begin{equation}
		\widehat{\mathcal{Q}}_{i}^{Bol} 
		\!=\! \iint_{\mathbb{R}^3\times\mathbb{S}^2} \! H(g^2-\delta_{ij}^{hk})
		g\widehat{\sigma}_{ijhk}(g,\widehat{\Omega}\cdot\widehat{\Omega}^\prime) \!
		\left[ \! \left( \! \dfrac{\mu_{ij}}{\mu_{hk}}\right)^{\!3} \!\! f_h(\mathbf{v}^\prime)f_k(\mathbf{w}^\prime)
		\!-\! f_i(\mathbf{v})f_j(\mathbf{w}) \! \right] \! d\mathbf{w}d\widehat{\Omega}^\prime\,,
		\label{Bolt_chem}
	\end{equation}
	where $H$ is the Heaviside function and $\mu_{ij}=m_im_j/(m_i+m_j)$ denotes the reduced mass {of the colliding pair}. 
	The quantity $\delta_{ij}^{hk}=2\Delta E/\mu_ij$ provides a proper threshold for the modulus of relative velocity to be overcome 
	in order to guarantee the {occurrence of the} exothermic reaction.
	
	As above, pre- and post-interaction velocities fulfill the conservation
	\begin{equation}
		\begin{aligned}
			m_i+m_j&=m_h+m_k\\
			m_i\mathbf{v}+m_j\mathbf{w}&=m_h\mathbf{v}^\prime+m_k\mathbf{w}^\prime\\
			\dfrac12m_iv^2+E_i + \dfrac12m_jw^2+E_j&=\dfrac12m_h(v^\prime)^2+E_h+\dfrac12m_k(w^\prime)^2+E_k\,,
		\end{aligned}
		\label{cons_ch}
	\end{equation}
	{Moreover, the quantity $\widehat{\sigma}_{ijhk}(g,\widehat{\Omega}\cdot\widehat{\Omega}^\prime)$} 
	denotes the {reactive} cross-section.
	{Here,} we assume {a chemical potential of} Maxwell {type} also for chemically reacting {interactions,}
	\begin{equation}
		\widehat{\lambda}_{ijhk}^{(0)}=\int_{\mathbb{S}^2} g\widehat{\sigma}_{ijhk}(g,\widehat{\Omega}\cdot\widehat{\Omega}^\prime)d\widehat{\Omega}^\prime=\text{constant}\,.
	\end{equation}
	and {symmetry relations similar to those in \eqref{symmetrycross} hold for reactive interactions.}
	
	\medskip
	
	As well known, collision equilibria {of the reactive mixture entail both mechanical and chemical equilibrium.
		They are defined by} distribution functions of Maxwell type, for which collision operators \eqref{Bolt_mech} and \eqref{Bolt_chem} 
	simultaneously vanish, {that is}
	
	\begin{equation}
		f_i^M(\mathbf{v}) 
		= n_i \, \mathcal{M}(\mathbf{v};\mathbf{u},T)
		= n_i\left(\dfrac{m_i}{2\pi T}\right)^\frac32\exp\left[-\dfrac{m_i}{2T}\left(\mathbf{v}-\mathbf{u}\right)^2\right]\,,
		\label{eq:MaxMx}
	\end{equation}
	{where, in addition,} species number densities fulfill the mass action law
	\begin{equation}
		\dfrac{n_1n_2}{n_3n_4}=\left(\dfrac{m_1m_2}{m_3m_4}\right)^\frac32\exp\left(\dfrac{\Delta E}{T}\right) .
		\label{MAL}
	\end{equation}
	Mechanically, the equilibrium states are described by equating species velocities and temperatures, 
	so that all constituents share a common velocity and a common temperature, which are those of the mixture.
	The mass action law \eqref{MAL} gives a characterization of such states from a chemical point of view
	and represents the chemical equilibrium condition of the chemical reaction \eqref{reaction}.
	
	\medskip
	
	In addition, in the space homogeneous {conditions}, 
	it is possible to prove the {global} stability of {such} equilibria by showing that {the function}
	\begin{equation}
		\mathcal{H}[\mathbf{f}]=\sum_{i=1}^4\int_{\mathbb{R}^3}f_i(\mathbf{v})\log \left(\dfrac{f_i(\mathbf{v})}{m_i^3}\right)d\mathbf{v}
		\label{eq:Hchem}
	\end{equation}
	is a Lyapunov functional {for the kinetic system \eqref{kin_eq} with the corresponding collisional operator $\mathcal{Q}_i^{Bol}$, 
		thus} providing, consequently, an entropy estimation.
	
	
	\subsection{A BGK model}
	\label{ssec:bgk}
	
	Due to the {complexity of the} integral form of the Boltzmann operators, 
	some alternative formulations are {highly} desirable. 
	The most common {approach} is the {so called} BGK {model, which replaces} 
	the Boltzmann terms by simpler relaxation operators. {These operators are defined in terms of suitable relaxation attractors
		involving free parameters that must be specified to ensure the mathematical and physical consistency of the model.}
	
	Despite their simplicity, these operators, being non-integral and linear with respect to the distribution functions, 
	effectively reproduce the main results of the Boltzmann description, 
	including the conservation laws of mass, momentum, and total energy, correct equilibria, and entropy inequality.
	
	It is important to note that the BGK model for a single gas can be extended to mixtures in several ways.
	In particular, the usual conservation laws provide few constraints when setting the free auxiliary parameters in the relaxation attractors, 
	and additional conditions are required.
	
	Among the models proposed in the literature, we consider, in this paper, the one proposed in 
	\cite{BBGSP} for inert mixtures. Here, we extend this model to a  reactive 
	{mixture} in the presence of the chemical reaction given in \eqref{reaction}. 
	The aim is to replicate the same structure of the Boltzmann collision operator, as a sum of many terms, 
	one for each type of interaction. 
	In this way, the resulting model is able to capture a broader range of mixture effects. 
	Accordingly, the BGK operator can be decomposed as
	\begin{equation}
		\mathcal{Q}_i^{BGK}
		= \widetilde{\mathcal{Q}}_i^{BGK}+\widehat{\mathcal{Q}}_i^{BGK}
		= \sum_{j=1}^4\widetilde{\mathcal{Q}}_{ij}^{BGK} + \widehat{\mathcal{Q}}_i^{BGK}\,,
	\end{equation}
	with mechanical terms $\widetilde{\mathcal{Q}}_{ij}^{BGK}$ and chemical terms $\widehat{\mathcal{Q}}_{i}^{BGK}$
	given by
	\begin{equation}
		\widetilde{\mathcal{Q}}_{ij}^{BGK}=\widetilde{\nu}_{ij}\left(\mathcal{M}_{ij}-f_i\right)\,,\quad i,j=1,2,3,4\,,
	\end{equation}
	and
	\begin{equation}
		\widehat{\mathcal{Q}}_{i}^{BGK}=\widehat{\nu}_{ij}^{hk}\left(\mathcal{M}_{i}-f_i\right)\,,\quad (i,j,h,k)\in\left\{(1,2,3,4),(2,1,4,3),(3,4,1,2),(4,3,2,1)\right\}\,,
		\label{Q_ch_BGK}
	\end{equation}
	respectively,
	where $\widetilde{\nu}_{ij}$ represents the frequency of mechanical collisions between $i-$th and $j-$th components, 
	and $\widehat{\nu}_{ij}^{hk}$ the frequency of reactive collisions obeying the chemical law \eqref{reaction}.
	Moreover, 
	$$
	\mathcal{M}_{ij} = \widetilde n_{ij} \, \mathbf M_i(\mathbf{v};\widetilde{\mathbf{u}}_{ij},\widetilde{T}_{ij})
	\qquad \mbox{and} \qquad
	\mathcal{M}_{i} = \widehat n_i \, \mathbf M_i(\mathbf{v};\widehat{\mathbf{u}}_{i},\widehat{T}_{i})
	$$
	are attractors depending on some fictitious parameters to be determined,
	namely
	$\widetilde{n}_{ij}$, $\widetilde{\mathbf{u}}_{ij}$, $\widetilde{T}_{ij}$, 
	and $\widehat{n}_i$, $\widehat{\mathbf{u}}_i$, $\widehat{T}_i$.
	
	
	\section{Computation of auxiliary parameters}
	\label{sec:par}
	
	In order to determine the auxiliary parameters in the Maxwellian attractors, we follow the procedure proposed in 
	\cite{BBGSP}.
	Hence, we assume that the exchange rates of momentum and energy for Boltzmann and BGK formulations coincide for any type of interaction.
	This approach assures that the BGK model is, by construction, consistent with the correct conservation laws
		and the correct exchange coefficients for the considered mixture. 
	
	For this purpose, we define
	\begin{equation}
		\begin{aligned}
			\widetilde{P}_{ij}^{(k)}=\int_{\mathbb{R}^3}\varphi_i^{(k)}(\mathbf{v})\widetilde{Q}_{ij}^{Bol}(\mathbf{v})d\mathbf{v}\quad&,\quad\widehat{P}_i^{(k)}=\int_{\mathbb{R}^3}\varphi_i^{(k)}(\mathbf{v})\widehat{Q}_{i}^{Bol}(\mathbf{v})d\mathbf{v}\,,\\
			\widetilde{\Pi}_{ij}^{(k)}=\int_{\mathbb{R}^3}\varphi_i^{(k)}(\mathbf{v})\widetilde{Q}_{ij}^{BGK}(\mathbf{v})d\mathbf{v}\quad&,\quad\widehat{\Pi}_i^{(k)}=\int_{\mathbb{R}^3}\varphi_i^{(k)}(\mathbf{v})\widehat{Q}_{i}^{BGK}(\mathbf{v})d\mathbf{v}\,,
		\end{aligned}
	\end{equation}
	where
	\begin{equation}
		\varphi_i^{(0)}(\mathbf{v})=m_i\,,\quad\varphi_i^{(1)}(\mathbf{v})=m_i\mathbf{v}\,,\quad\varphi_i^{(2)}(\mathbf{v})=\dfrac12m_iv^2+E_i\,.
	\end{equation}
	
	
	\subsection{Parameters in mechanical terms}
	\label{ssec:parm}
	
	As anticipated above, we assume that
	\begin{equation}
		\widetilde{\Pi}_{ij}^{(k)}=\widetilde{P}_{ij}^{(k)},\quad i,j=1,2,3,4\,,\quad k=0,1,2\,.
		\label{ass_mech}
	\end{equation}
	To compute the production terms coming from Boltzmann formulation, we consider the following weak form of Boltzmann operators
	\begin{equation}
		\widetilde{P}_{ij}^{(k)}=\int_{\mathbb{R}^3\times \mathbb{R}^3 \times\mathbb{S}^2}g\widetilde{\sigma}_{ij}(g,\widehat{\Omega}\cdot\widehat{\Omega}^\prime)\left(\varphi_i^{(k)}(\mathbf{v}^\prime)-\varphi_i^{(k)}(\mathbf{v})\right)f_i(\mathbf{v})f_j(\mathbf{w}) d\mathbf{v}d\mathbf{w}d\widehat{\Omega}^\prime\,,
	\end{equation}
	and, under the hypothesis of Maxwell molecule potential, we can compute explicitly
	\begin{equation}
		\begin{aligned}
			\widetilde{P}_{ij}^{(0)}&=0\\
			\widetilde{P}_{ij}^{(1)}&=\widetilde{\lambda}_{ij}^{(0)}\mu_{ij}n_in_j(\mathbf{u}_j-\mathbf{u}_i)\\
			\widetilde{P}_{ij}^{(2)}&=\widetilde{\lambda}_{ij}^{(0)}\alpha_{ij}\alpha_{ji}n_in_j\left[(m_i\mathbf{u}_i+m_j\mathbf{u}_j)\cdot(\mathbf{u}_j-\mathbf{u}_i)+3(T_j-T_i)\right]\,,
		\end{aligned}
	\end{equation}
	where we notice that
	\begin{equation}
		\int_{\mathbb{S}^2}\widehat{\Omega}^\prime\widetilde{\sigma}_{ij}(g,\widehat{\Omega}\cdot\widehat{\Omega}^\prime)d\widehat{\Omega}^\prime=\mathbf{0}
	\end{equation}
	by parity arguments, and we made use of the relation between pre- and post-collision velocities
	\begin{equation}
		\begin{aligned}
			\mathbf{v}^\prime&=\alpha_{ij}\mathbf{v}+\alpha_{ji}\mathbf{w}+\alpha_{ji}g^\prime\widehat{\Omega}^\prime\\
			\mathbf{w}^\prime&=\alpha_{ij}\mathbf{v}+\alpha_{ji}\mathbf{w}-\alpha_{ij}g^\prime\widehat{\Omega}^\prime\,,
		\end{aligned}
	\end{equation}
	$\alpha_{ij}=m_i/(m_i+m_j)$ denoting the mass ratio.
	
	For what concerns the production terms from the BGK model, we can easily obtain that
	\begin{equation}
		\begin{aligned}
			\widetilde{\Pi}_{ij}^{(0)}&=\widetilde{\nu}_{ij}m_i(\widetilde{n}_{ij}-n_i)\\
			\widetilde{\Pi}_{ij}^{(1)}&=\widetilde{\nu}_{ij}m_i(\widetilde{n}_{ij}\widetilde{\mathbf{u}}_{ij}-n_i\mathbf{u}_i)\\
			\widetilde{\Pi}_{ij}^{(2)}&=\widetilde{\nu}_{ij}\left[\dfrac12m_i\left(\widetilde{n}_{ij}\widetilde{\mathbf{u}}_{ij}^2-n_i\mathbf{u}_i^2\right)+\dfrac32\left(\widetilde{n}_{ij}\widetilde{T}_{ij}-n_iT_i\right)+E_i(\widetilde{n}_{ij}-n_i)\right]\,,
		\end{aligned}
	\end{equation}
	and hence, by assumption \eqref{ass_mech}, we get
	\begin{equation}
		\begin{aligned}
			\widetilde{n}_{ij}&=n_i\\
			\widetilde{\mathbf{u}}_{ij}&=(1-a_{ij})\mathbf{u}_i+a_{ij}\mathbf{u}_j\\
			\widetilde{T}_{ij}&=(1-b_{ij})T_i+b_{ij}T_j+\gamma_{ij}|\mathbf{u}_i-\mathbf{u}_j|^2
		\end{aligned}
		\label{fict_mech}
	\end{equation}
	with
	\begin{equation}
		a_{ij}=\dfrac{\widetilde{\lambda}_{ij}^{(0)}}{\widetilde{\nu}_{ij}}\alpha_{ji}n_j\,,\quad b_{ij}=2a_{ij}\alpha_{ij}\,,\quad\gamma_{ij}=\dfrac13m_ia_{ij}(2\alpha_{ji}-a_{ij})\,.
	\end{equation}
	The positivity of auxiliary temperatures is guaranteed under the condition $\widetilde{\nu}_{ij}\ge\dfrac12\widetilde{\lambda}^{(0)}_{ij}n_j$.
	
	Further details can be found in \cite{BBGSP}.
	
	
	\subsection{Parameters in chemical terms}
	\label{ssec:parch}
	
	As done for mechanical contributions, we impose that the exchange rates of mass, momentum, and total energy due to chemical interactions are the same for Boltzmann and BGK operators, i.e.
	\begin{equation}
		\widehat{\Pi}_{i}^{(k)}=\widehat{P}_{i}^{(k)},\quad  i=1,2,3,4\,,\quad k=0,1,2\,.
		\label{ass_chem}
	\end{equation}
	Production terms in the BGK model can be computed explicitly, while the corresponding contributions in the Boltzmann formulation are approximated by considering a quasi-equilibrium configuration \cite{Kremer-PandolfiBianchi-Soares-PoF2006}.
	We consider
	\begin{equation}
		f_i(\mathbf{v})\simeq n_i\left(\dfrac{m_i}{2\pi T}\right)^{3/2}\exp\left[-\dfrac{m_i}{2T}|\mathbf{v}-\mathbf{u}|^2\right]\mathcal{E}_i(\mathbf{v})\,,i=1,2,3,4\,,
		\label{f_i_app}
	\end{equation}
	where
	\begin{equation}
		\mathcal{E}_i(\mathbf{v})=\mathcal{A}_i+\mathcal{B}_i(\mathbf{v}-\mathbf{u})\cdot(\mathbf{u}_i-\mathbf{u})+\mathcal{C}_i|\mathbf{v}-\mathbf{u}|^2
		\label{expr_E_i}
	\end{equation}
	and
	\begin{equation}
		\mathcal{A}_i=1-\dfrac32\,\dfrac{T_i-T}{T}\,,\quad\mathcal{B}_i=\dfrac{m_i}{2T}\,,\quad\mathcal{C}_i=\dfrac{m_i}{2T}\,\dfrac{T_i-T}{T}\,.
	\end{equation}
	We get
	\begin{equation}
		\begin{aligned}
			\widehat{n}_{i}&=n_i+\dfrac{\widehat{P}_{i}^{(0)}}{m_i\widehat{\nu}_{ij}^{hk}}\\
			\widehat{\mathbf{u}}_{i}&=\dfrac{1}{\widehat{n}_i}\left(n_i\mathbf{u}_i+\dfrac{\widehat{P}_{i}^{(1)}}{m_i\widehat{\nu}_{ij}^{hk}}\right)\\
			\widehat{T}_{i}&=\dfrac{1}{\widehat{n}_i}\left\{n_iT_i+\dfrac23\left[\dfrac{\widehat{P}_{i}^{(2)}}{\widehat{\nu}_{ij}^{hk}}-\dfrac12m_i\left(\widehat{n}_i\widehat{\mathbf{u}}_i^2-n_i\mathbf{u}_i^2\right)-E_i\left(\widehat{n}_i-n_i\right)\right]\right\}\,,
		\end{aligned}
	\end{equation}
	where production terms $\widehat{P}_{i}^{(0)}$, $\widehat{P}_{i}^{(1)}$, $\widehat{P}_{i}^{(2)}$ are computed in detail in Appendix  \ref{App}.
	
	We can notice that 
	\begin{equation}
		\dfrac{\widehat{P}_1^{(0)}}{m_1}=\dfrac{\widehat{P}_2^{(0)}}{m_2}=-\dfrac{\widehat{P}_3^{(0)}}{m_3}=-\dfrac{\widehat{P}_4^{(0)}}{m_4}
	\end{equation}
	and hence the partial and total conservation of mass follows; moreover, also global momentum and energy preservation are correctly reproduced, i.e. it holds
	\begin{equation}
		\begin{aligned}
			&\sum_{i=1}^4\widehat{P}_i^{(1)}=\boldsymbol{0}\\
			&\sum_{i=1}\widehat{P}_i^{(2)}+\lambda_i\dfrac{\widehat{P}_i^{(0)}}{m_i}\Delta E=0\,,\text{ for any fixed }i=1,2,3,4\,,
		\end{aligned}
	\end{equation}
	where $\lambda_1=\lambda_2=1=-\lambda_3=-\lambda_4$ are the stoichiometric coefficients.\\
	Since $\mathcal{M}_{i}=\widehat{n}_i\mathbf M_i(\mathbf{v};\widehat{\mathbf{u}}_{i},\widehat{T}_{i})$ is a distribution function for each species $i=1,2,3,4$, the Maxwellian must be positive and measurable, and hence we must require that
	\begin{equation}
		\widehat{n}_i>0 \quad \text{and} \quad \widehat{T}_i>0 .
		\label{pos_chem}
	\end{equation}
	The condition \eqref{pos_chem} is fulfilled at least in the quasi-equilibrium configuration, 
	since the production terms $\widehat{P}_{i}$ deduced from Boltzmann operators are sufficiently small.
	
	
	\section{Entropy properties and trend to equilibrium}
	\label{sec:Htheo}
	
	One important property of the model is its consistency with the entropy growth and 
		the correct definition of the equilibrium states for the reactive mixture. 
		This result is crucial when studying hydrodynamic limits and convergence towards equilibrium.
		
		In this section, we first establish an $\cal H$-theorem for the proposed BGK model and define the equilibrium states compatible with 
		the entropy production.  
		Then, we present some numerical simulations that illustrate the approach to equilibrium in light of the $\cal H$-theorem.

	
	\subsection{Entropy growth}
	\label{ssec:Htheo}
	
	Regrettably, proving rigorously an $\mathcal{H}-$theorem for the model proposed above,
	in its rather general formulation, seems
	unfeasible.
	An entropy estimate can be established only under specific assumptions, 
	more precisely, assuming the equalization of fictitious species velocities and temperatures in the chemical BGK contributions, 
	i.e.
	\begin{equation}
		{\widehat{\mathbf{u}}{_i}} = \widehat{\mathbf{u}},
		\qquad
		\widehat{T}_i=\widehat{T}\,.
		\label{equal_cond}
	\end{equation}
	Moreover, we assume that auxiliary densities fulfill the following algebraic constraint
	accounting for the mass action law \eqref{MAL},
	\begin{equation}
		\dfrac{\widehat{n}_1\widehat{n}_2}{\widehat{n}_3\widehat{n}_4}
		=  \left(\dfrac{m_1m_2}{m_3m_4}\right)^\frac32\exp\left(\dfrac{\Delta E}{\widehat{T}}\right) .
		\label{MAL_chem}
	\end{equation}
	\\
	{
		Note that such assumptions \eqref{equal_cond} and \eqref{MAL_chem} are distinctive hypotheses
		when proving an $\mathcal{H}$-theorem for a BGK model for reactive mixtures, as can be seen, for instance, in the model proposed in \cite{GS2004} and then reformulated in \cite{BGS2008}, or in the papers \cite{BPS2014} and \cite{Brull-Schneider-CMS2014}.
		In our work, thus, the general shape of attractors allows us to reproduce the correct conservation laws, whereas we need the extra assumptions \eqref{equal_cond} and \eqref{MAL_chem}
		to prove an $\mathcal{H}$-theorem. 
		More in detail,}
	it is possible to prove that the $\mathcal{H}$-functional for chemically reactive mixtures,
	already introduced in \eqref{eq:Hchem} for the Boltzmann system, that is
	\begin{equation}
		\mathcal{H}\left[f_i\right]=\sum_{i=1}^4\int_{\mathbb{R}^3}f_i(\mathbf{v})\log\left(\dfrac{f_i(\mathbf{v})}{m_i^3}\right) d\mathbf{v} ,
		\label{eq:HchemK}
	\end{equation}
	is a Lyapunov functional for our BGK model proposed in Subsection \ref{ssec:bgk}. 
	The convergence to equilibrium can be proved by combining the techniques already performed in \cite{BBGSP} and in \cite{Brull-Schneider-CMS2014}. For this reason, we state the results and summarize the main steps of the proof.
			We address the reader to the cited works and references therein for further details. 
	
	\begin{theorem}
		{(\,$\cal H$-theorem)}
		Let us assume {conditions} \eqref{equal_cond} and \eqref{MAL_chem} {on the auxiliary parameters of chemical terms}. 
		Under space homogeneous conditions, for all measurable distribution functions $f_i\ge 0$, $i=1,2,3,4$, 
		we have that
		\begin{equation}
			\dfrac{d\mathcal{H}}{d t}\le 0\,, \quad \text{for all} \;\; t\ge 0\,.
		\end{equation}
		\label{th:HH}
	\end{theorem}
	\textit{Proof.} Under space homogeneous conditions, one has
	\begin{equation}
		\begin{aligned}
			\dfrac{d\mathcal{H}}{dt}&=\sum_{i=1}^4\int_{\mathbb{R}^3}\dfrac{\partial f_i}{\partial t}\log\left(\dfrac{f_i}{m_i^3}\right)d\mathbf{v}+\sum_{i=1}^4\int_{\mathbb{R}^3}f_i\dfrac{m_i^3}{f_i}\dfrac{1}{m_i^3}\dfrac{\partial f_i}{\partial t}d\mathbf{v}\\
			&=\sum_{i=1}^4\sum_{j=1}^4\int_{\mathbb{R}^3}\widetilde{\nu}_{ij}\left(\mathcal{M}_{ij}-f_i\right)\log\left(\dfrac{f_i}{m_i^3}\right)d\mathbf{v}+\sum_{i=1}^4\sum_{j=1}^4\widetilde{\nu}_{ij}\int_{\mathbb{R}^3}\left(\mathcal{M}_{ij}-f_i\right)d\mathbf{v}\\
			&+\sum_{i=1}^4\int_{\mathbb{R}^3}\widehat{\nu}_{ij}^{hk}\left(\mathcal{M}_{i}-f_i\right)\log\left(\dfrac{f_i}{m_i^3}\right)d\mathbf{v}+\sum_{i=1}^4\widehat{\nu}_{ij}^{hk} \int_{\mathbb{R}^3}\left(\mathcal{M}_{i}-f_i\right)d\mathbf{v}\,,
		\end{aligned}
		\label{eq:Hs1}
	\end{equation}
	where, {taking into account the assumptions on the auxiliary parameters, the chemical attractor is given by}
	\begin{equation}
		\mathcal{M}_{i}=\widehat{n}_i\mathbf M_i(\mathbf{v};\widehat{\mathbf{u}},\widehat{T})
		= \widehat{n}_i\left(\dfrac{m_i}{2\pi \widehat{T}}\right)^\frac32\exp
		\left[-\dfrac{m_i}{2\widehat{T}}\left(\mathbf{v}-\widehat{\mathbf{u}}\right)^2\right] .
	\end{equation}
	{Concerning the attractors, we} observe that
	\begin{equation}
		\int_{\mathbb{R}^3}\left(\mathcal{M}_{ij}-f_i\right)d\mathbf{v}=\widetilde{n}_{ij}-n_i=n_i-n_i=0
		\label{eq:M1chem}
	\end{equation}
	and
	\begin{equation}
		\sum_{i=1}^4\widehat{\nu}_{ij}^{hk} \int_{\mathbb{R}^3}\left(\mathcal{M}_{i}-f_i\right)d\mathbf{v} 
		=\sum_{i=1}^4\widehat{\nu}_{ij}^{hk}(\widehat{n}_i-n_i)=\sum_{i=1}^4\dfrac{\widehat{P}_i^{(0)}}{m_i}=0\,.
	\end{equation}
	Therefore, {coming back to \eqref{eq:Hs1}, we obtain}
	\begin{equation}
		\begin{aligned}
			\dfrac{d\mathcal{H}}{dt}
			&=\sum_{i=1}^4\sum_{j=1}^4 \! \int_{\mathbb{R}^3}\widetilde{\nu}_{ij} \!
			\left(\mathcal{M}_{ij}-f_i\right)\log \! \left(\dfrac{f_i}{m_i^3}\right) \! d\mathbf{v}
			+\sum_{i=1}^4 \!\int_{\mathbb{R}^3}\widehat{\nu}_{ij}^{hk}
			\left(\mathcal{M}_{i}-f_i\right)\log\! \left(\dfrac{f_i}{m_i^3}\right) \! d\mathbf{v}
			\\
			&=\sum_{i=1}^4\sum_{j=1}^4 \! \int_{\mathbb{R}^3}\widetilde{\nu}_{ij} \!
			\left(\mathcal{M}_{ij}-f_i\right) \log \! \left(\dfrac{f_i}{m_i^3}\right) \! d\mathbf{v}
			+\sum_{i=1}^4\int_{\mathbb{R}^3}\widehat{\nu}_{ij}^{hk}
			\left(\mathcal{M}_{i}-f_i\right)\log \! \left(\dfrac{f_i}{\mathcal{M}_{i}}\right) \! d\mathbf{v}\\
			&+\sum_{i=1}^4\int_{\mathbb{R}^3}\widehat{\nu}_{ij}^{hk}\left(\mathcal{M}_{i}-f_i\right)\log\left(\dfrac{\mathcal{M}_{i}}{m_i^3}\right)d\mathbf{v}
		\end{aligned}
		\label{H_sum}
	\end{equation}
	
	\bigskip
	
	\noindent
	{Now, we treat separately the three contributions on the right-hand side of expression \eqref{H_sum}.}
	
	\medskip
	
	\noindent
	The first contribution can be dealt with as done in \cite{BBGSP}
	{and it can be shown that
		\begin{equation}
			\sum_{i=1}^4\sum_{j=1}^4 \int_{\mathbb{R}^3}\widetilde{\nu}_{ij}\left(\mathcal{M}_{ij}-f_i\right)\log\left(\dfrac{f_i}{m_i^3}\right)d\mathbf{v}
			\leq 0.
			\label{eq:C1}
		\end{equation}
	}

	\medskip
	
	\noindent
	Coming back to \eqref{H_sum}, {and considering the second contribution, 
		it is easy to prove that} 
	{
		\begin{equation}
			\int_{\mathbb{R}^3}\widehat{\nu}_{ij}^{hk}\left(\mathcal{M}_{i}-f_i\right)\log\left(\dfrac{f_i}{\mathcal{M}_{i}}\right)d\mathbf{v} 
			\leq 0.
			\label{eq:C2}
		\end{equation}
		More precisely, we have}
	\begin{equation}
		\int_{\mathbb{R}^3}\widehat{\nu}_{ij}^{hk}\left(\mathcal{M}_{i}-f_i\right)\log\left(\dfrac{f_i}{\mathcal{M}_{i}}\right)d\mathbf{v}=\int_{\mathbb{R}^3}\widehat{\nu}_{ij}^{hk}\mathcal{M}_{i}\left(1-\dfrac{f_i}{\mathcal{M}_{i}}\right)\log\left(\dfrac{f_i}{\mathcal{M}_{i}}\right)d\mathbf{v}\le 0\,,
	\end{equation}
	by means of usual convexity arguments of the function $(1-x)\log x$. 
	
	\medskip
	
	\noindent
	Finally, {considering the third contribution} in \eqref{H_sum}, 
	{and it vanishes \cite{Brull-Schneider-CMS2014}, that is
		\begin{equation}
			\begin{aligned}
				\sum_{i=1}^4\int_{\mathbb{R}^3}\widehat{\nu}_{ij}^{hk}
				\left(\mathcal{M}_{i}-f_i\right)\log\left(\dfrac{\mathcal{M}_{i}} {m_i^3}\right)d\mathbf{v}
				&=  \dfrac{\widehat{P}_1^{(0)}}{m_1}\left\{\log\left[\dfrac{\widehat{n}_1\widehat{n}_2}{\widehat{n}_3\widehat{n}_4}\left(\dfrac{m_3m_4}{m_1m_2}\right)^{3/2}\right]-\dfrac{\Delta E}{\widehat{T}}\right\} \\
				&=  0,
			\end{aligned}\label{eq:C3}
		\end{equation}	
	{where assumption} \eqref{MAL_chem} {associated to the mass action law is used
		to justify the vanishing of the above integral term.}
	This completes the proof.
	\hfill$\square$
	
	\bigskip
	
	
	As a consequence of Theorem \ref{th:HH}, we can prove that the equilibrium states for our BGK model
	are defined by the Maxwellian distribution functions given by \eqref{eq:MaxMx}, \eqref{MAL},
	that assure the equilibrium of the reactive mixture.
	
	\medskip
	
	
	\begin{theorem}
		\label{th:HHb}
		Let assume conditions \eqref{equal_cond} and \eqref{MAL_chem} on the auxiliary para\-me\-ters of chemical terms. 
		Under space homogeneous conditions, for all measurable distribution functions $f_i\ge 0$, $i=1,2,3,4$, 
		we have that
		\begin{equation}
			\frac{d{\cal H}}{dt}(t) = 0 \qquad \mbox{iff} \qquad f_i = f_i^M, \ \ \mbox{for}  \quad i=1,2,3,4.
			\label{eq:Hp2}
		\end{equation}
	\end{theorem}
	\textit{Proof.}
	In Theorem \ref{th:HH}, we have {proven} that 
	\begin{equation}
		\dfrac{d\mathcal{H}}{dt}
		= \sum_{i=1}^4 \! \sum_{j=1}^4\int_{\mathbb{R}^3} \! \widetilde{\nu}_{ij}
		\left(\mathcal{M}_{ij} \!-\! f_i\right) \log \! \left(\dfrac{f_i}{m_i^3}\right) \! d\mathbf{v}
		+ \sum_{i=1}^4\int_{\mathbb{R}^3} \! \widehat{\nu}_{ij}^{hk}
		\left(\mathcal{M}_{i} \!-\! f_i\right) \log \! \left(\dfrac{f_i}{\mathcal{M}_{i}}\right) \! d\mathbf{v} ,
		\label{H_sum2}
	\end{equation}
	where
	\begin{equation}
		\sum_{i=1}^4\sum_{j=1}^4\int_{\mathbb{R}^3}\widetilde{\nu}_{ij}\left(\mathcal{M}_{ij}-f_i\right)\log\left(\dfrac{f_i}{m_i^3}\right) \! d\mathbf{v} \leq 0
		\label{H_sum3}
	\end{equation}
	and
	\begin{equation}
		\sum_{i=1}^4\int_{\mathbb{R}^3}\widehat{\nu}_{ij}^{hk}\left(\mathcal{M}_{i}-f_i\right)\log\left(\dfrac{f_i}{\mathcal{M}_{i}}\right) \! d\mathbf{v} \leq 0
		\label{H_sum4}
	\end{equation}
	since conditions \eqref{eq:C1}, \eqref{eq:C2} and \eqref{eq:C3} hold.
	
	\medskip
	
	Therefore, the time derivative of the $\cal H$-function vanishes if and only if 
	{relations \eqref{H_sum3} and \eqref{H_sum4} hold as equalities. 
		However, as discussed in \cite{BBGSP}, the equality in \eqref{H_sum3} implies that the auxiliary velocities 
		$\widetilde{\mathbf{u}}_i$ and temperatures $\widetilde{T}_i$ 
		share a common value $\widetilde{\mathbf{u}}$ and $\widetilde{T}$, respectively, and hence 
		$f_i = n_i \, \mathbf M_i(\mathbf{v};\widetilde{\mathbf{u}},\widetilde{T})$, for $i=1,2,3,4$.
		Thus, all constituents of the mixture share a common velocity $\widetilde{\mathbf{u}}$
		and a common temperature $\widetilde T$,
		assuring mechanical equilibrium of the mixture.
		Now, plugging in the above distributions $f_i$ in \eqref{H_sum4},
		the equality implies that also 
		$n_i \, \mathbf M_i(\mathbf{v};\widetilde{\mathbf{u}},\widetilde{T}) = \mathcal{M}_{i}$.
		Under condition \eqref{equal_cond}, this gives
		$\mathbf M_i(\mathbf{v};\widetilde{\mathbf{u}},\widetilde{T})
		\!=\! \mathbf M_i(\mathbf{v};\widehat{\mathbf{u}},\widehat{T})$
		and therefore
		$\widetilde{\mathbf{u}} \!=\! \widehat{\mathbf{u}}$ and $\widetilde{T} \!=\! \widehat{T}$.
		Now, condition \eqref{MAL_chem} implies that distributions
		$f_i $ also verify the chemical equilibrium condition and therefore
		$f_i = f_i^M $, for $i=1,2,3,4$.
		This completes the proof.
		\hfill$\square$}
	
	
	\subsection{Numerical insight}
	\label{ssec:numerics}
	
	In this subsection, we perform numerical simulations to investigate the equilibrium approach of a reacting mixture under spatially homogeneous conditions, whose relaxation to equilibrium is guaranteed by the previous $\cal H$-theorem under suitable assumptions. We rely on the same scenario provided in \cite{GS2004}, and we carry out our analysis with the additional assumption of isotropic distribution functions, i.e.
	$f_i(\mathbf v)=f_i(v)$, that leads to $\mathbf{u}_i=\mathbf{u}=\mathbf{0}$.
	
	In order to represent a specific scenario, we rely on the reversible chemical reaction already modeled in \cite{BT2} which involves
	nitrosyl chloride $ClNO$ (65,46 g/mol), nitrogen dioxide $NO_2$ (46.01 g/mol),
	chloro nitride $ClNO_2$ (81.46 g/mol) and nitric oxide $NO$ (30.01 g/mol)
	\begin{equation}\label{reaz}
		NO_2+ClNO\leftrightarrows ClNO_2+NO.
	\end{equation}
	
	Even if constituents are polyatomic, we neglect here the internal structure of the reactants and we consider them as monatomic species, leaving a more detailed description of polyatomic gases for future development of the present work.
	
	The explicit computation of production terms and auxiliary parameters  (whose details are reported in Appendix \ref{App})
	allows us to investigate the behavior of macroscopic fields, which are evaluated using an appropriate trapezoidal quadrature scheme. The resulting system of partial differential equations is discretized in terms of the microscopic velocity modulus, and the resulting discrete system of ordinary differential equations is then solved numerically, adopting a standard Runge-Kutta-based.

	We set the initial configuration coherently with the hypotheses of
	Theorems \ref{th:HH}-\ref{th:HHb}, i.e. we start from a close-to-equilibrium configuration. Specifically, we assume that each component of the mixture initially has a Maxwellian-shaped distribution function, expressed as
	\begin{equation}
		f_i^0(\mathbf{v}) 
		= n_i^0\left(\dfrac{m_i}{2\pi T_i^0}\right)^{\!\frac32} \exp\left[-\dfrac{m_i}{2T^0_i}\left({v}\right)^2\right]\,.
		\label{eq:MaxMx}
	\end{equation}
	
	We choose masses having ratios comparable to the  ones of the gases involved in the reaction \eqref{reaz}: 
	\begin{equation}\label{masse}
		m_1=2.18,\quad m_2=1.53,\quad m_3=1,\quad m_4=2.72.
	\end{equation}
	As internal energies of the four components, we take the following ones
	\begin{equation}\label{energie} 
		E_1=7.5\times10^{-4},\quad E_2=1\times10^{-3},\quad E_3=7\times10^{-4},\quad E_4=1.2\times10^{-3}.
	\end{equation}
	Then, we fix, as initial densities and temperatures, the following
	\begin{equation}\label{densIn1}
		n_1^0=1	,\quad n_2^0=1.2,\quad n_3^0=1.4,\quad n_4^0=1.3,
	\end{equation}
	\begin{equation}\label{tempIn1}   
		T_1^0=4\times 10^{-2},\quad T_2^0=4.3\times 10^{-2},\quad T_3^0=3.7\times 10^{-2},\quad T_4^0=3.5\times 10^{-2}.
	\end{equation}
	To compute the auxiliary fields for BGK operators, we take the momenta of the cross-section and collision frequencies as
	\begin{equation}
		\tilde{\lambda}_{ij}=0.001,\quad i,j=1,2,3,4,\quad
		\tilde{\nu}_{ij}=\left(\begin{array}{cccc}
			3 & 4 & 1 & 4 \\
			4 & 3 & 4 & 6 \\
			1 & 4 & 3 & 2 \\
			4 & 6 & 2 & 4
		\end{array}\right),
	\end{equation}
	\begin{equation}
		\hat{\nu}_{ij}^{hk}=1,\quad (i,j),(h,k)\in\big\{(1,2),(3,4),(2,1),(4,3)\big\}, i\neq h,i\neq k.
	\end{equation}
	
	Our formulation allows us to observe, separately, the relaxation of auxiliary temperatures for each species accounting for the mechanical interactions with every other species and for chemical reactions, separately. In particular, the trend depicted in  Figure \ref{aux_temp} shows that the relaxation of chemical auxiliary temperatures $\hat{T}_i$ to $T_i$ takes place later than the convergence of mechanical fictitious ones $\tilde{T}_{ij}$. Even if the profile of $\hat{T}_i/T_i$ crosses the value $1$ at a very early stage, during the transient.
	\begin{figure}[ht!]
		\includegraphics[trim={0 0 14cm 0},clip,scale=0.37]{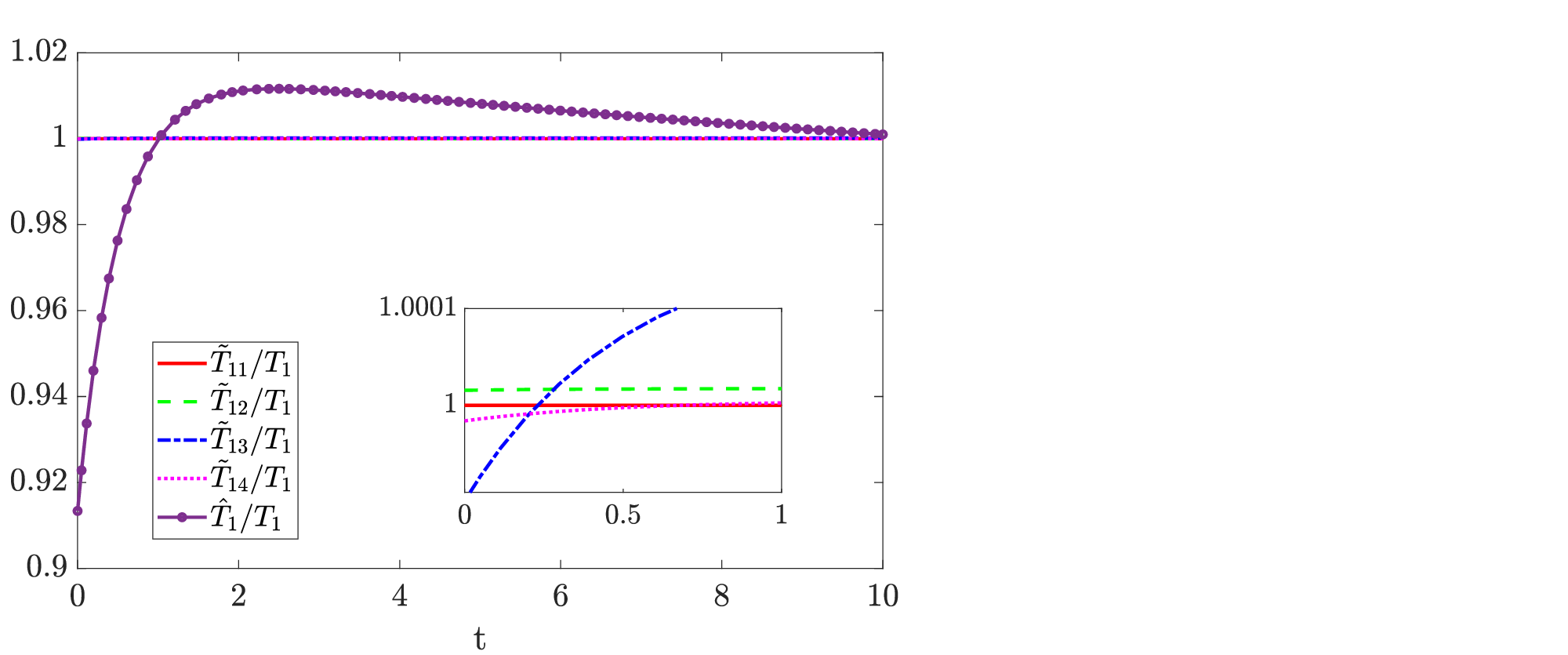}
		\includegraphics[trim={0 0 14cm 0},clip,scale=0.37]{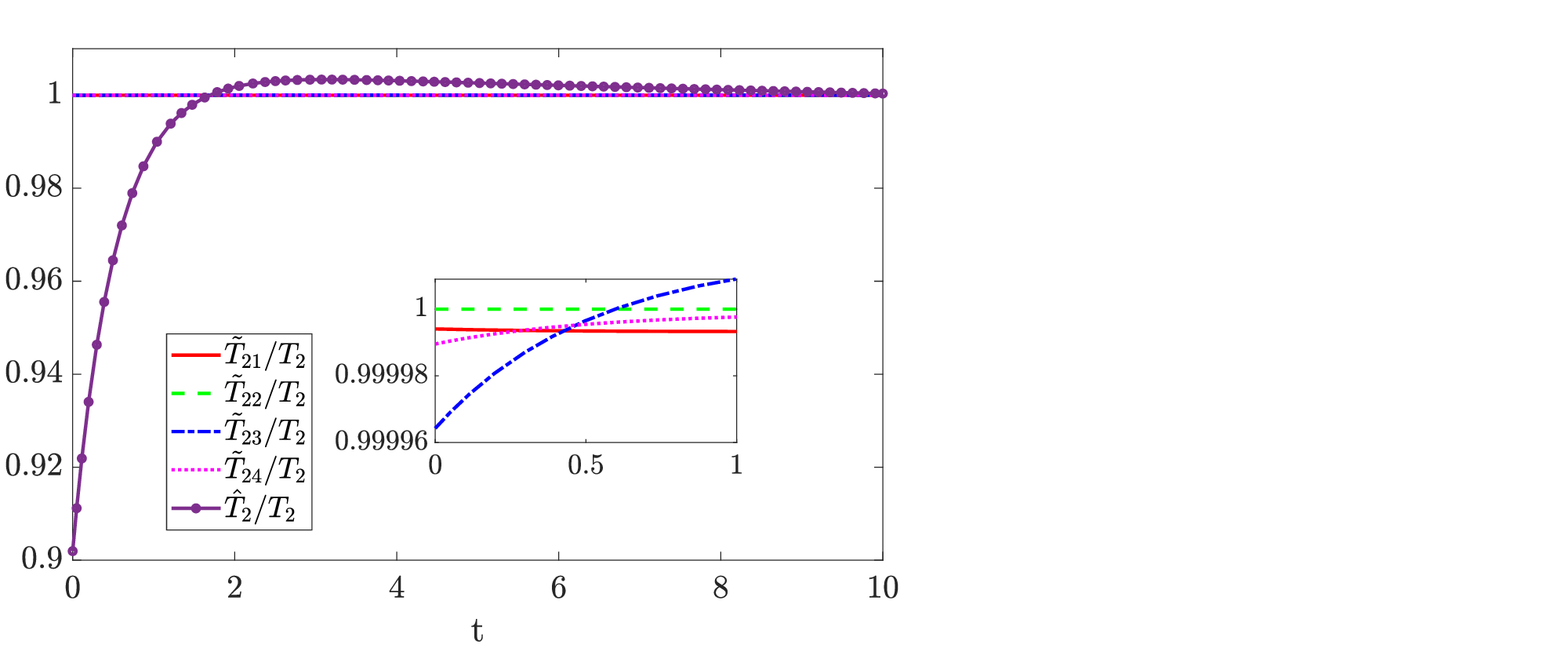}\\
		\includegraphics[trim={0 0 14cm 0},clip,scale=0.37]{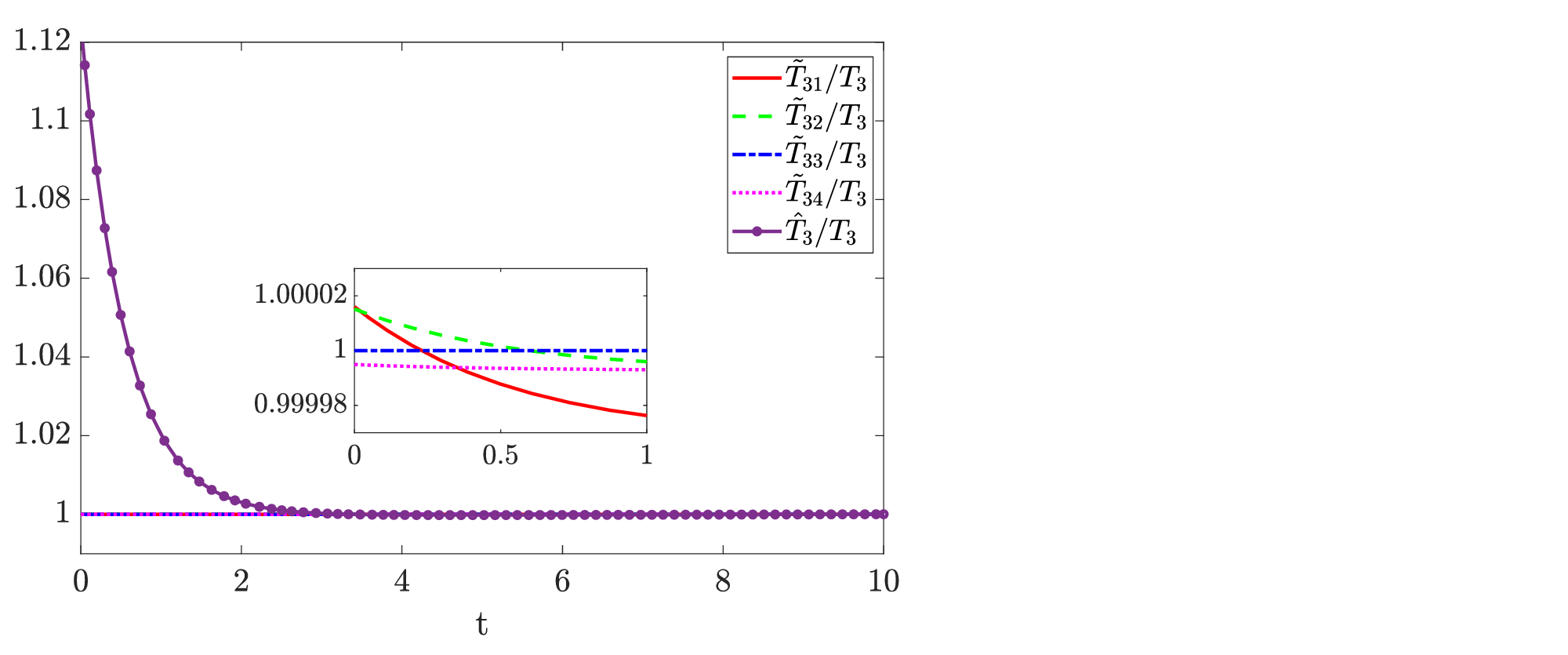}
		\includegraphics[trim={0 0 14cm 0},clip,scale=0.37]{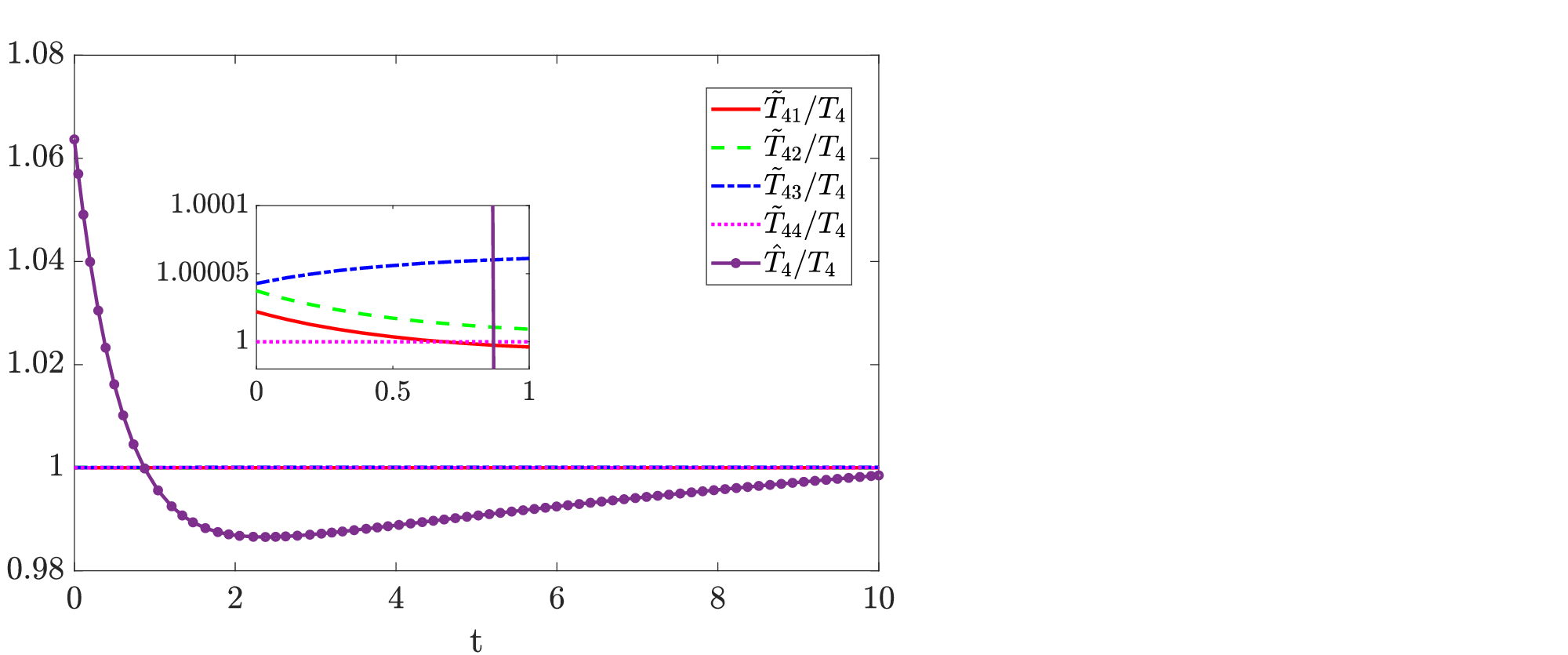}
		\caption{Auxiliary temperatures in mechanical and chemical BGK terms, rescaled with respect to species temperature.}
		\label{aux_temp}
	\end{figure}
	In figure \ref{Htheo}, instead, we compute the $H$ functional defined in \eqref{eq:HchemK}. We observe that starting from a close-to-equilibrium configuration, the $H$ functional has a monotone trend, as inferred in Theorems \ref{th:HH} - \ref{th:HHb}, even if the assumption \eqref{equal_cond} is not imposed. Thus we can state that, under suitable conditions, the model is consistent also if rigorous proof cannot be provided.
	
	\begin{figure}[ht!]
		\includegraphics[scale=0.4]{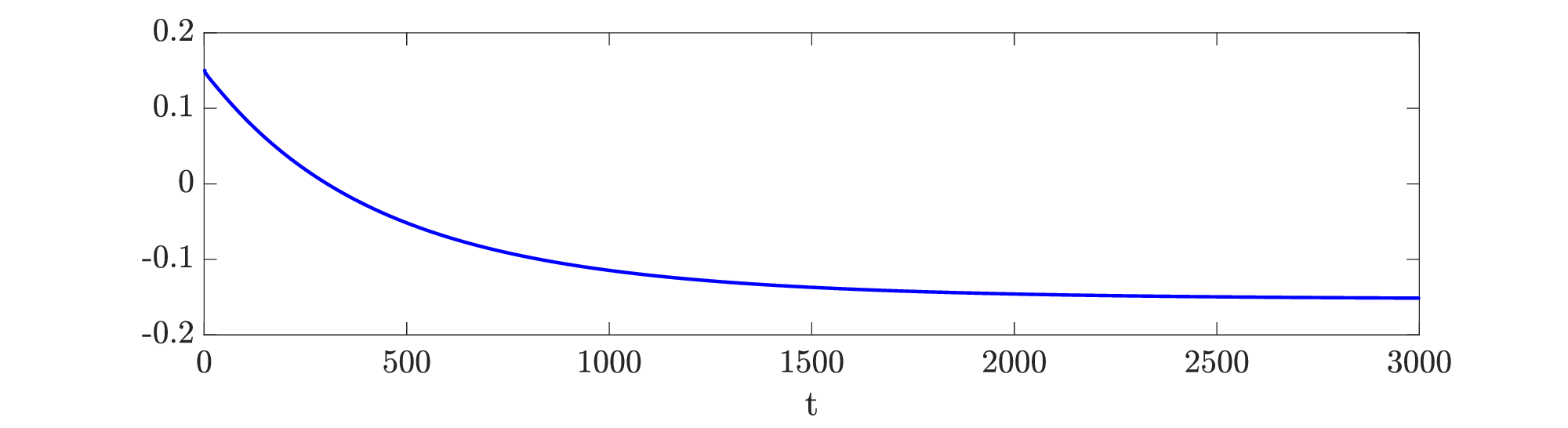}
		\caption{H-functional given in \eqref{eq:HchemK}.}
		\label{Htheo}
	\end{figure}

	\bigskip
	
	
	\section{Conclusions}
	\label{sec:conc}
	We have proposed a BGK-type kinetic model for a reactive mixture of four monatomic gases undergoing a bimolecular reversible chemical reaction. 
	Although the classical Boltzmann description is able to take {into} account {the} intermolecular potential {of} the microscopic {collisional}, a BGK formulation {is} highly desirable, being more manageable from {both the} analytical and {the} numerical point of view.
	
	The model {proposed} in this {work} extends  the description given in \cite{BBGSP} for an inert mixture, 
		where the collision operator is the sum of BGK terms, one for each pair of interacting components.
		Moreover, this formulation overcomes the limitation of \cite{Brull-Schneider-CMS2014}, where the chemical operators share the same mean velocity and temperature, and thus they do not account for the specific features of the constituents.
	We have assumed that the exchange rates of momentum and energy {coincide} for Boltzmann and BGK formulations,
	both in mechanical terms, as done in \cite{BBGSP}, and {in} chemical contributions. 
	We point out that the Boltzmann production terms for the chemical operator cannot be computed explicitly, 
	unless a proper approximation of the distribution function is assumed.
	In this paper, we consider the perturbation of the Maxwellian distribution used in \cite{Kremer-PandolfiBianchi-Soares-PoF2006}, 
	corresponding to a quasi-equilibrium configuration.
	
	The new model proposed here has the great advantage of modeling separately the effects of a chemical reaction and the ones of the mechanical interactions. Moreover, as observed in \cite{bisi2021macroscopic}, the mechanical collision operator splits in a sum of binary terms, which takes into account the individual role of any pair of interacting gas species {and, thus retains the mixture effects in the collisional dynamics}. 
	Therefore, the BGK equations of this model mimic the structure of the Boltzmann formulation and hence provide the possibility to derive macroscopic evolution equations for the main fields in different hydrodynamic {chemical} regimes, according to the dominant collisional phenomenon.
	
	Additionally, since the model explicitly separates mechanical and chemical terms,
		it can be applied to investigate transport coefficients in different evolution regimes,
		including fast and slow chemical reactions.
		We plan to address this problem in the near future, leveraging the general potential of BGK models and, 
		in particular, and the capabilities of the proposed model.
	
	
	As usual in the framework {of the reactive mixtures}, two distinctive features emerge. 
	At first, the number density of each constituent is not preserved, since particles can change their nature when they interact at the chemical level.
	{T}his implies that auxiliary number densities appearing in the Maxwellian attractors for chemical contributions result in nonlinear functions of masses, species concentration, global auxiliary temperature, {and} energy threshold, whose shape derives from model assumptions. 
	The second feature regards the characterization of equilibria from a chemical point of view, 
	which allows us to reproduce correctly the mass action law of chemistry.
	
	Due to these characteristics, some technical difficulties may arise when proving the consistency properties of the model, such as the positivity of number densities and temperatures, as well as the entropy estimate. As regards the latter issue, a proper H-theorem has been proved by assuming in the BGK terms for chemical interactions the equalization of the auxiliary velocities and temperatures and an additional algebraic constraint relating auxiliary densities to global temperature.
	
	Our future projects will involve further development of the ideas in this paper to extend a {recent} mixed Boltzmann-BGK model to reactive gaseous mixtures. Moreover, it would be interesting to test the proposed model in space-dependent problems, 
	such as the analysis of the one-dimensional shock wave structure.

	\vspace*{1cm}
	
	
	\section*{Acknowledgments}
	The work of authors G.M. and R.T. was performed in the frame of activities supported by Italian National Group of Mathematical Physics (GNFM-INdAM) 
	and by the University of Parma (Italy).
	R.T. is a post-doc fellow supported by the National Institute of Advanced Mathematics (INdAM), Italy. 
	
	\smallskip
	
	\noindent
	The research work of R.T. and A.J.S. was carried out in the frame of activities sponsored by the Cost Action CA18232, 
	by the Portuguese Projects UIDB/00013/2020 (\url{https://doi.org/10.54499/UIDB/00013/2020}), 
	UIDP/00013/2020 of CMAT-UM (\url{https://doi.org/10.54499/UIDP/00013/2020}), 
	and by the Portuguese national funds (OE), through the Project FCT/MCTES  PTDC/03091/2022, 
	``Mathematical Modelling of Multi-scale Control Systems: applications to human diseases -- CoSysM3''
	(\url{https://doi.org/10.54499/2022.03091.PTDC}).
	
	\smallskip
	
	\noindent
	The work of R.T.  was carried out in the frame of activities sponsored by University of Parma through the action 
	Bando di Ateneo 2022 per la ricerca co-funded by MUR-Italian Ministry of Universities and 
	Research - D.M. 737/2021 - PNR - PNRR - NextGenerationEU, 
	Project ``Collective and self-organised dynamics: kinetic and network approaches''.
	
	\smallskip
	
	\noindent
	The research work of A.J.S. was carried out in the frame of activities sponsored by Pessoa Project Ref.2021.09255.CBM.
	
	%
	
	\bigskip\bigskip
	

	
	
	\appendix
	
	\section{Appendix -- computation of integrals}
	\label{App}
	
	\medskip
	
	{In this appendix, we include supplementary material, 
		showing the details of the computations involved in the determination of
		the production terms of mass, momentum, and energy.
		The computations are quite lengthy and somewhat intricate.
		We choose to provide more details, focusing solely on the production terms of one constituent.
		For the other constituents, the computations are rather similar.
		
		More specifically, w}e calculate the source terms $\widehat{P}_{1}^{(k)}$, $k=0,1,2$, for the first constituent.
	{By} using relations \eqref{cons_ch}, such terms can be written as
	\begin{equation}
		\widehat{P}_1^{(k)}=\dfrac{(m_1m_2)^{3/2}}{(2\pi T)^3}\left[\left(\dfrac{m_1m_2}{m_3m_4}\right)^{3/2}\exp\left(\dfrac{\Delta E}{T}\right)n_3n_4\mathcal{R}_{1+}^{k}-n_1n_2\mathcal{R}_{1-}^{(k)}\right]\,,
	\end{equation}
	where
	\begin{equation}
		\begin{aligned}
			\mathcal{R}_{1+}^{(k)}&=\int_{\mathbb{R}^3\times\mathbb{R}^3\times\mathbb{S}^2}H(g^2-\delta_{12}^{34})g\widehat{\sigma}_{1234}(g,\widehat{\Omega}\cdot\widehat{\Omega}^\prime)\varphi_1^{(k)}(\mathbf{v})\\
			&\hspace{1.5cm}\times\mathcal{E}_3(\mathbf{v}^\prime)\mathcal{E}_4(\mathbf{w^\prime})\exp\left[-\dfrac{m_1}{2T}|\mathbf{v}-\mathbf{u}|^2-\dfrac{m_2}{2T}|\mathbf{w}-\mathbf{u}|^2\right]d\mathbf{v}d\mathbf{w}d\widehat{\Omega}^\prime\\
			\mathcal{R}_{1-}^{(k)}&=\int_{\mathbb{R}^3\times\mathbb{R}^3\times\mathbb{S}^2}H(g^2-\delta_{12}^{34})g\widehat{\sigma}_{1234}(g,\widehat{\Omega}\cdot\widehat{\Omega}^\prime)\varphi_1^{(k)}(\mathbf{v})\\
			&\hspace{1.5cm}\times\mathcal{E}_1(\mathbf{v})\mathcal{E}_2(\mathbf{w})\exp\left[-\dfrac{m_1}{2T}|\mathbf{v}-\mathbf{u}|^2-\dfrac{m_2}{2T}|\mathbf{w}-\mathbf{u}|^2\right]d\mathbf{v}d\mathbf{w}d\widehat{\Omega}^\prime\,.
		\end{aligned}
	\end{equation}
	Similar expressions for other {constituents} can be recovered through equivalent forms of the weak formulation of the collision operator \cite{rossani1999note}.

	
	\subsection{Some preliminaries}
	\label{ssec:prelim}
	
	It is convenient to introduce {the following} change of variable with unit jacobian,
	\begin{equation}
		(\mathbf{v},\mathbf{w})\,\longrightarrow\,(\mathbf{v}^*,\mathbf{w}^*)\,\longrightarrow\,(\mathbf{G}^*,\mathbf{g}^*)\,,
	\end{equation}
	where
	\begin{equation}
		\mathbf{v}^*=\mathbf{v}-\mathbf{u}\,,\quad\mathbf{w}^*=\mathbf{w}-\mathbf{u}\,,\quad\mathbf{G}^*=\alpha_{12}\mathbf{v}^*+\alpha_{21}\mathbf{w}^*\,,\quad\mathbf{g}^*=\mathbf{v}^*-\mathbf{w}^*=\mathbf{g}\,.
	\end{equation}
	Therefore, we can rewrite
	\begin{equation}
		-\dfrac{m_1}{2T}|\mathbf{v}-\mathbf{u}|^2-\dfrac{m_2}{2T}|\mathbf{w}-\mathbf{u}|^2=-\dfrac{m_1+m_2}{2T}(G^*)^2-\dfrac{\mu_{12}}{2T}g^2\,,
	\end{equation}
	and some of the following integrals intervene
	\begin{equation}
		\begin{aligned}
			I_{\alpha\beta\gamma}^\ell&=\int_{\mathbb{R}^3}(G_1^*)^{2\alpha}(G_2^*)^{2\beta}(G_3^*)^{2\gamma}(G^*)^{2\ell}\exp\left(-\dfrac{m_1+m_2}{2T}(G^*)^2\right)d\mathbf{G}^*\\
			Y_{\alpha\beta\gamma}^\ell&=\int_{\mathbb{R}^3}H(g^2-\delta_{12}^{34})g_1^{2\alpha}g_2^{2\beta}g_3^{2\gamma}g^{2\ell}\exp\left(-\dfrac{\mu_12}{2T}g^2\right)d\mathbf{g}
		\end{aligned}
	\end{equation}
	when computing $\mathcal{R}_{1\pm}^{(k)}$. Both types of integrals can be computed explicitly in terms of complete and incomplete Gamma function
	\begin{equation}
		\Gamma(\alpha)=\int_0^{+\infty}\tau^{\alpha-1}e^{-\tau}d\tau\,,\quad\Gamma(\alpha,\beta)=\int_\beta^{+\infty}\tau^{\alpha-1}e^{-\tau}d\tau
	\end{equation}

	
	\subsection{Production of mass}
	\label{ssec:mass}
	
	We have to compute $\mathcal{R}_{1\pm}^{(0)}$, by using the test function $\varphi_1^{(0)}(\mathbf{v})=m_1$.
	From \eqref{expr_E_i}, we {have}
	\begin{equation}
		\begin{aligned}
			\mathcal{E}_3(\mathbf{v}^\prime)\mathcal{E}_4(\mathbf{w^\prime})&=\left[\mathcal{A}_3+\mathcal{B}_3\mathbf{v}^{\prime*}\cdot\mathbf{u}_3^*+\mathcal{C}_3(v^{\prime*})^2\right]\left[\mathcal{A}_4+\mathcal{B}_4\mathbf{w}^{\prime*}\cdot\mathbf{u}_4^*+\mathcal{C}_4(w^{\prime*})^2\right]\,,
		\end{aligned}
		\label{E3E4}
	\end{equation}
	where
	\begin{equation}
		\begin{aligned}
			\mathbf{v}^{\prime*}&=\mathbf{G}^*+\alpha_{43}\sqrt{\dfrac{\mu_{12}}{\mu_{34}}\left(g^2-\delta_{12}^{34}\right)}\widehat{\Omega}^\prime\,,\quad\mathbf{w}^{\prime*}=\mathbf{G}^*-\alpha_{34}\sqrt{\dfrac{\mu_{12}}{\mu_{34}}\left(g^2-\delta_{12}^{34}\right)}\widehat{\Omega}^\prime\,,
		\end{aligned}
	\end{equation}
	and hence the term $\mathcal{R}_{1+}^{(0)}$ can be rewritten as sum of several integrals in terms of $\mathbf{G}^*$ and $\mathbf{g}$; by parity arguments, some of these terms vanish and, consequently, $\mathcal{R}_{1+}^{(0)}$ reduces to
	\begin{equation}
		\begin{aligned}
			\mathcal{R}_{1+}^{(0)} &=\mathcal{J}_{1+}^{(0)}+\mathcal{J}_{2+}^{(0)}+\mathcal{J}_{3+}^{(0)}+\mathcal{J}_{4+}^{(0)}+\mathcal{J}_{5+}^{(0)}\,,
		\end{aligned}
	\end{equation}
	where
	\begin{equation}
		\begin{aligned}
			J_{1+}^{(0)}&=m_1\mathcal{A}_3\mathcal{A}_4 \int_{\mathbb{R}^3\times\mathbb{R}^3\times\mathbb{S}^2}H(g^2-\delta_{12}^{34})g\widehat{\sigma}_{1234}(g,\widehat{\Omega}\cdot\widehat{\Omega}^\prime) \\ &\hspace{2.5cm}\times\exp\left[-\dfrac{m_1+m_2}{2T}(G^*)^2-\dfrac{\mu_{12}}{2T}g^2\right]d\mathbf{G}^*d\mathbf{g}d\widehat{\Omega}^\prime\\
			&=m_1\mathcal{A}_3\mathcal{A}_4\widehat{\lambda}_{1234}^{(0)} I_{000}^0Y_{000}^0
		\end{aligned}
	\end{equation}
	\begin{equation}
		\begin{aligned}
			J_{2+}^{(0)}&=m_1\mathcal{A}_3\mathcal{C}_4 \int_{\mathbb{R}^3\times\mathbb{R}^3\times\mathbb{S}^2}H(g^2-\delta_{12}^{34})g\widehat{\sigma}_{1234}(g,\widehat{\Omega}\cdot\widehat{\Omega}^\prime)\\
			&\hspace{1.5cm}\times\left[(G^*)^2-2\alpha_{34}\sqrt{\dfrac{\mu_{12}}{\mu_{34}}\left(g^2-\delta_{12}^{34}\right)}\mathbf{G}^*\cdot\widehat{\Omega}^\prime+\alpha_{34}^2\dfrac{\mu_{12}}{\mu_{34}}\left(g^2-\delta_{12}^{34}\right)\right] \\
			&\hspace{1.5cm}\times\exp\left[-\dfrac{m_1+m_2}{2T}(G^*)^2-\dfrac{\mu_{12}}{2T}g^2\right]d\mathbf{G}^*d\mathbf{g}d\widehat{\Omega}^\prime\\
			&=m_1\mathcal{A}_3\mathcal{C}_4\widehat{\lambda}_{1234}^{(0)} \left[I_{000}^1Y_{000}^0+\alpha_{34}^2\dfrac{\mu_{12}}{\mu_{34}}I_{000}^0\left(Y_{000}^1-\delta_{12}^{34}Y_{000}^0\right)\right]
		\end{aligned}
	\end{equation}
	\begin{equation}
		\begin{aligned}
			J_{3+}^{(0)}&=m_1\mathcal{A}_4\mathcal{C}_3 \int_{\mathbb{R}^3\times\mathbb{R}^3\times\mathbb{S}^2}H(g^2-\delta_{12}^{34})g\widehat{\sigma}_{1234}(g,\widehat{\Omega}\cdot\widehat{\Omega}^\prime)\\
			&\hspace{1.5cm}\times\left[(G^*)^2+2\alpha_{43}\sqrt{\dfrac{\mu_{12}}{\mu_{34}}\left(g^2-\delta_{12}^{34}\right)}\mathbf{G}^*\cdot\widehat{\Omega}^\prime+\alpha_{43}^2\dfrac{\mu_{12}}{\mu_{34}}\left(g^2-\delta_{12}^{34}\right)\right] \\
			&\hspace{1.5cm}\times\exp\left[-\dfrac{m_1+m_2}{2T}(G^*)^2-\dfrac{\mu_{12}}{2T}g^2\right]d\mathbf{G}^*d\mathbf{g}d\widehat{\Omega}^\prime\\
			&=m_1\mathcal{A}_4\mathcal{C}_3\widehat{\lambda}_{1234}^{(0)} \left[I_{000}^1Y_{000}^0+\alpha_{43}^2\dfrac{\mu_{12}}{\mu_{34}}I_{000}^0\left(Y_{000}^1-\delta_{12}^{34}Y_{000}^0\right)\right]
		\end{aligned}
	\end{equation}
	\begin{equation}
		\begin{aligned}
			J_{4+}^{(0)}&=m_1\mathcal{B}_3\mathcal{B}_4 \int_{\mathbb{R}^3\times\mathbb{R}^3\times\mathbb{S}^2}H(g^2-\delta_{12}^{34})g\widehat{\sigma}_{1234}(g,\widehat{\Omega}\cdot\widehat{\Omega}^\prime) \\
			&\hspace{1.5cm}\times\left[\left(\mathbf{G}^*+\alpha_{43}\sqrt{\dfrac{\mu_{12}}{\mu_{34}}\left(g^2-\delta_{12}^{34}\right)}\widehat{\Omega}^\prime\right)\cdot\mathbf{u}_3^*\right] \\
			&\hspace{1.5cm}\times\left[\left(\mathbf{G}^*-\alpha_{34}\sqrt{\dfrac{\mu_{12}}{\mu_{34}}\left(g^2-\delta_{12}^{34}\right)}\widehat{\Omega}^\prime\right)\cdot\mathbf{u}_4^*\right]  \\
			&\hspace{1.5cm}\times\exp\left[-\dfrac{m_1+m_2}{2T}(G^*)^2-\dfrac{\mu_{12}}{2T}g^2\right]d\mathbf{G}^*d\mathbf{g}d\widehat{\Omega}^\prime\\
			&=m_1\mathcal{B}_3\mathcal{B}_4 \int_{\mathbb{R}^3\times\mathbb{R}^3\times\mathbb{S}^2}H(g^2-\delta_{12}^{34})g\widehat{\sigma}_{1234}(g,\widehat{\Omega}\cdot\widehat{\Omega}^\prime) \\
			&\hspace{1.5cm}\times\left[\sum_{i=1}^3(G_i^*)^2u_{3i}^*u_{4i}^*-\alpha_{34}\alpha_{43}\dfrac{\mu_{12}}{\mu_{34}}\left(g^2-\delta_{12}^{34}\right)\sum_{i,j=1}^3\widehat{\Omega}_i^\prime\widehat{\Omega}_j^\prime u_{3i}^*u_{4j}^*\right] \\
			&\hspace{1.5cm}\times\exp\left[-\dfrac{m_1+m_2}{2T}(G^*)^2-\dfrac{\mu_{12}}{2T}g^2\right]d\mathbf{G}^*d\mathbf{g}d\widehat{\Omega}^\prime\\
			&=m_1\mathcal{B}_3\mathcal{B}_4\left[\widehat{\lambda}_{1234}^{(0)}Y_{000}^0\left(I_{100}^0u_{31}^*u_{41}^*+I_{010}^0u_{32}^*u_{42}^*+I_{001}^0u_{33}^*u_{43}^*\right)\right.\\
			&\hspace{3.5cm}-\alpha_{34}\alpha_{43}\dfrac{\mu_{12}}{\mu_{34}}\left.I_{000}^0\left(Y_{000}^1-\delta_{12}^{34}Y_{000}^0\right)\widehat{\boldsymbol{\Lambda}}_{1234}^{(2)}:(\mathbf{u}_3^* \!\otimes\! \mathbf{u}_4^*)\right]\\
			&=m_1\mathcal{B}_3\mathcal{B}_4 \! \left[\widehat{\lambda}_{1234}^{(0)}I_{100}^0Y_{000}^0\mathbf{u}_3^*\cdot\mathbf{u}_4^* \!-\! \alpha_{12}\alpha_{21}I_{000}^0\left(Y_{000}^1 \!-\! \delta_{12}^{34}Y_{000}^0\right)\widehat{\boldsymbol{\Lambda}}_{1234}^{(2)}:(\mathbf{u}_3^*\otimes\mathbf{u}_4^*)\right]
		\end{aligned}
	\end{equation}
	with
	\begin{equation}
		I^0_{100}=I^0_{010}=I^0_{001}
	\end{equation}
	and
	\begin{equation}
		\widehat{\boldsymbol{\Lambda}}_{1234}^{(2)}=\int_{\mathbb{S}^2} \left(\widehat{\Omega}^\prime\otimes\widehat{\Omega}^\prime\right) g\widehat{\sigma}_{1234}(g,\widehat{\Omega}\cdot\widehat{\Omega}^\prime)d\widehat{\Omega}^\prime=\text{constant}\,.
	\end{equation}
	\begin{equation}
		\begin{aligned}
			J_{5+}^{(0)}&=m_1\mathcal{C}_3\mathcal{C}_4 \int_{\mathbb{R}^3\times\mathbb{R}^3\times\mathbb{S}^2}H(g^2-\delta_{12}^{34})g\widehat{\sigma}_{1234}(g,\widehat{\Omega}\cdot\widehat{\Omega}^\prime) \\
			&\hspace{1.5cm}\times\left[(G^*)^2+2\alpha_{43}\sqrt{\dfrac{\mu_{12}}{\mu_{34}}\left(g^2-\delta_{12}^{34}\right)}\mathbf{G}^*\cdot\widehat{\Omega}^\prime+\alpha_{43}^2\dfrac{\mu_{12}}{\mu_{34}}\left(g^2-\delta_{12}^{34}\right)\right] \\
			&\hspace{1.5cm}\times\left[(G^*)^2-2\alpha_{34}\sqrt{\dfrac{\mu_{12}}{\mu_{34}}\left(g^2-\delta_{12}^{34}\right)}\mathbf{G}^*\cdot\widehat{\Omega}^\prime+\alpha_{34}^2\dfrac{\mu_{12}}{\mu_{34}}\left(g^2-\delta_{12}^{34}\right)\right]  \\
			&\hspace{1.5cm}\times\exp\left[-\dfrac{m_1+m_2}{2T}(G^*)^2-\dfrac{\mu_{12}}{2T}g^2\right]d\mathbf{G}^*d\mathbf{g}d\widehat{\Omega}^\prime\\
			&=m_1\mathcal{C}_3\mathcal{C}_4\int_{\mathbb{R}^3\times\mathbb{R}^3\times\mathbb{S}^2}H(g^2-\delta_{12}^{34})g\widehat{\sigma}_{1234}(g,\widehat{\Omega}\cdot\widehat{\Omega}^\prime) \\
			&\hspace{1.5cm}\times\left[(G^*)^4-4\alpha_{34}\alpha_{43}\dfrac{\mu_{12}}{\mu_{34}}\left(g^2-\delta_{12}^{34}\right)\sum_{ij=1}^3G_i^*G_j^*\widehat{\Omega}_i^\prime\widehat{\Omega}_j^\prime\right.\\
			&\hspace{2.5cm}\left.+\alpha_{34}^2\alpha_{43}^2 \! \left(\dfrac{\mu_{12}}{\mu_{34}}\right)^{\!2} \! 
			\left(g^2-\delta_{12}^{34}\right)^{\!2}+\left(\alpha_{34}^2+\alpha_{43}^2\right)\dfrac{\mu_{12}}{\mu_{34}}(G^*)^2\left(g^2-\delta_{12}^{34}\right)\right] \\
			&\hspace{1.5cm}\times\exp\left[-\dfrac{m_1+m_2}{2T}(G^*)^2-\dfrac{\mu_{12}}{2T}g^2\right]d\mathbf{G}^*d\mathbf{g}d\widehat{\Omega}^\prime\\
			&=m_1\mathcal{C}_3\mathcal{C}_4 \left\{\widehat{\lambda}_{1234}^{(0)}I_{000}^2Y_{000}^0-4\alpha_{12}\alpha_{21}I_{100}^0\left(Y_{000}^1-\delta_{12}^{34}Y_{000}^0\right)\text{tr}(\widehat{\boldsymbol{\Lambda}}_{1234}^{(0)})\right.\\
			&\hspace{1.5cm}+\widehat{\lambda}_{1234}^{(0)}\alpha_{12}^2\alpha_{21}^2I_{000}^0\left[Y_{000}^2-2\delta_{12}^{34}Y_{000}^1+\left(\delta_{12}^{34}\right)^2Y_{000}^0\right]\\
			&\hspace{1.5cm}\left.+\widehat{\lambda}_{1234}^{(0)}\left(\alpha_{34}^2+\alpha_{43}^2\right)\dfrac{\mu_{12}}{\mu_{34}}I_{000}^1\left(Y_{000}^1-\delta_{12}^{34}Y_{000}^0\right)\right\}
		\end{aligned}
	\end{equation}
	Analogously, for the loss term, we have that
	\begin{equation}
		\begin{aligned}
			\mathcal{E}_1(\mathbf{v})\mathcal{E}_2(\mathbf{w})&=\left[\mathcal{A}_1+\mathcal{B}_1\mathbf{v}^{*}\cdot\mathbf{u}_1^*+\mathcal{C}_1(v^{*})^2\right]\left[\mathcal{A}_2+\mathcal{B}_2\mathbf{w}^{*}\cdot\mathbf{u}_2^*+\mathcal{C}_2(w^{*})^2\right]\,,
		\end{aligned}
		\label{E1E2}
	\end{equation}
	and $\mathcal{R}_{1-}^{(0)}$ reduces to
	\begin{equation}
		\begin{aligned}
			\mathcal{R}_{1-}^{(0)} &=\mathcal{J}_{1-}^{(0)}+\mathcal{J}_{2-}^{(0)}+\mathcal{J}_{3-}^{(0)}+\mathcal{J}_{4-}^{(0)}+\mathcal{J}_{5-}^{(0)}\,,
		\end{aligned}
	\end{equation}
	where
	\begin{equation}
		\begin{aligned}
			J_{1-}^{(0)}&=m_1\mathcal{A}_1\mathcal{A}_2 \int_{\mathbb{R}^3\times\mathbb{R}^3\times\mathbb{S}^2}H(g^2-\delta_{12}^{34})g\widehat{\sigma}_{1234}(g,\widehat{\Omega}\cdot\widehat{\Omega}^\prime) \\ &\hspace{2.5cm}\times\exp\left[-\dfrac{m_1+m_2}{2T}(G^*)^2-\dfrac{\mu_{12}}{2T}g^2\right]d\mathbf{G}^*d\mathbf{g}d\widehat{\Omega}^\prime\\
			&=m_1\mathcal{A}_1\mathcal{A}_2\widehat{\lambda}_{1234}^{(0)} I_{000}^0Y_{000}^0
		\end{aligned}
	\end{equation}
	\begin{equation}
		\begin{aligned}
			J_{2-}^{(0)}&=m_1\mathcal{A}_1\mathcal{C}_2 \int_{\mathbb{R}^3\times\mathbb{R}^3\times\mathbb{S}^2}H(g^2-\delta_{12}^{34})g\widehat{\sigma}_{1234}(g,\widehat{\Omega}\cdot\widehat{\Omega}^\prime)  \\
			&\hspace{1.5cm}\times\left[(G^*)^2-2\alpha_{12}\mathbf{G}^*\cdot\mathbf{g}+\alpha_{12}^2g^2\right]\exp\left[-\dfrac{m_1+m_2}{2T}(G^*)^2-\dfrac{\mu_{12}}{2T}g^2\right]d\mathbf{G}^*d\mathbf{g}d\widehat{\Omega}^\prime\\
			&=m_1\mathcal{A}_1\mathcal{C}_2\widehat{\lambda}_{1234}^{(0)} \left(I_{000}^1Y_{000}^0+\alpha_{12}^2I_{000}^0Y_{000}^1\right)
		\end{aligned}
	\end{equation}
	\begin{equation}
		\begin{aligned}
			J_{3-}^{(0)}&=m_1\mathcal{A}_2\mathcal{C}_1 \! \int_{\mathbb{R}^3\times\mathbb{R}^3\times\mathbb{S}^2}H(g^2-\delta_{12}^{34})g\widehat{\sigma}_{1234}(g,\widehat{\Omega}\cdot\widehat{\Omega}^\prime) \\
			&\hspace{1.0cm}\times \! \left[(G^*)^2 \!+\! 2\alpha_{21}\mathbf{G}^*\cdot\mathbf{g} \!+\! \alpha_{21}^2g^2\right] \!
			\exp \! \left[ \! -\dfrac{m_1\!+\!m_2}{2T}(G^*)^2-\dfrac{\mu_{12}}{2T}g^2 \! \right] \! d\mathbf{G}^*d\mathbf{g}d\widehat{\Omega}^\prime\\
			&=m_1\mathcal{A}_2\mathcal{C}_1\widehat{\lambda}_{1234}^{(0)} \left(I_{000}^1Y_{000}^0+\alpha_{21}^2I_{000}^0Y_{000}^1\right)
		\end{aligned}
	\end{equation}
	\begin{equation}
		\begin{aligned}
			J_{4-}^{(0)}&=m_1\mathcal{B}_1\mathcal{B}_2 \int_{\mathbb{R}^3\times\mathbb{R}^3\times\mathbb{S}^2}H(g^2-\delta_{12}^{34})g\widehat{\sigma}_{1234}(g,\widehat{\Omega}\cdot\widehat{\Omega}^\prime)\left[\left(\mathbf{G}^*+\alpha_{21}\mathbf{g}\right)\cdot\mathbf{u}_1^*\right] \\
			&\hspace{1.5cm}\times\left[\left(\mathbf{G}^*-\alpha_{12}\mathbf{g}\right)\cdot\mathbf{u}_2^*\right]\exp\left[-\dfrac{m_1+m_2}{2T}(G^*)^2-\dfrac{\mu_{12}}{2T}g^2\right]d\mathbf{G}^*d\mathbf{g}d\widehat{\Omega}^\prime\\
			&=m_1\mathcal{B}_1\mathcal{B}_2 \int_{\mathbb{R}^3\times\mathbb{R}^3\times\mathbb{S}^2}H(g^2-\delta_{12}^{34})g\widehat{\sigma}_{1234}(g,\widehat{\Omega}\cdot\widehat{\Omega}^\prime) \\
			&\hspace{1.5cm}\times\left[\sum_{i=1}^3(G_i^*)^2u_{1i}^*u_{2i}^*-\alpha_{12}\alpha_{21}\sum_{i=1}^3g_i^2u_{1i}^*u_{2i}^*\right] \\
			&\hspace{1.5cm}\times\exp\left[-\dfrac{m_1+m_2}{2T}(G^*)^2-\dfrac{\mu_{12}}{2T}g^2\right]d\mathbf{G}^*d\mathbf{g}d\widehat{\Omega}^\prime\\        &=m_1\mathcal{B}_1\mathcal{B}_2\widehat{\lambda}_{1234}^{(0)}\left(I_{100}^0Y_{000}^0-\alpha_{12}\alpha_{21}I_{000}^0Y_{000}^1\right)\mathbf{u}_1^*\cdot\mathbf{u}_2^*
		\end{aligned}
	\end{equation}
	with
	\begin{equation}
		Y_{100}^0=Y_{010}^0=Y_{001}^0\,.
	\end{equation}
	\begin{equation}
		\begin{aligned}
			J_{5-}^{(0)}&=m_1\mathcal{C}_1\mathcal{C}_2 \int_{\mathbb{R}^3\times\mathbb{R}^3\times\mathbb{S}^2}H(g^2-\delta_{12}^{34})g\widehat{\sigma}_{1234}(g,\widehat{\Omega}\cdot\widehat{\Omega}^\prime)\left[(G^*)^2+2\alpha_{21}\mathbf{G}^*\cdot\mathbf{g}+\alpha_{21}^2g^2\right] \\
			&\hspace{1.0cm}\times \! \left[(G^*)^2-2\alpha_{12}\mathbf{G}^*\cdot\mathbf{g} \!+\! \alpha_{12}^2g^2\right] \!
			\exp \! \left[-\dfrac{m_1\!+\! m_2}{2T}(G^*)^2 \!-\!  \dfrac{\mu_{12}}{2T}g^2\right]d\mathbf{G}^*d\mathbf{g}d\widehat{\Omega}^\prime\\
			&=m_1\mathcal{C}_1\mathcal{C}_2\int_{\mathbb{R}^3\times\mathbb{R}^3\times\mathbb{S}^2}H(g^2-\delta_{12}^{34})g\widehat{\sigma}_{1234}(g,\widehat{\Omega}\cdot\widehat{\Omega}^\prime) \\
			&\hspace{1.0cm}\times\left[(G^*)^4-4\alpha_{12}\alpha_{21}\sum_{i=1}^3(G_i^*)^2g_i^2+\alpha_{12}^2\alpha_{21}^2g^4+\left(\alpha_{12}^2+\alpha_{21}^2\right)(G^*)^2g^2\right] \\
			&\hspace{1.0cm}\times\exp\left[-\dfrac{m_1+m_2}{2T}(G^*)^2-\dfrac{\mu_{12}}{2T}g^2\right]d\mathbf{G}^*d\mathbf{g}d\widehat{\Omega}^\prime\\
			&=m_1 \mathcal{C}_1 \mathcal{C}_2  \widehat{\lambda}_{1234}^{(0)} \!\!
			\left[I_{000}^2 \! Y_{000}^0 \!\!-\!\! 12\alpha_{12}\alpha_{21}I_{100}^0 \! Y_{100}^0 \!\!+\!\! \alpha_{12}^2\alpha_{21}^2I_{000}^0 \! Y_{000}^2 \!\!+\!\!
			\left(\alpha_{12}^2 \!\!+\!\! \alpha_{21}^2\right)I_{000}^1Y_{000}^1\right]
		\end{aligned}
	\end{equation}
	
	
	\subsection{Production of momentum}
	\label{ssec:mom}
	
	We have to compute $\mathcal{R}_{1\pm}^{(1)}$, by using the test function $\varphi_1^{(0)}(\mathbf{v})=m_1\mathbf{v}^*$.\\
	More precisely, we have that $\mathcal{R}_{1+}^{(1)}$ is a vector and its first component $\left(\mathcal{R}_{1+}^{(1)}\right)_1$ reads as
	\begin{equation}
		\begin{aligned}
			\left(\mathcal{R}_{1+}^{(1)}\right)_1 &=\left(\mathcal{J}_{1+}^{(1)}\right)_1+\left(\mathcal{J}_{2+}^{(1)}\right)_1+\left(\mathcal{J}_{3+}^{(1)}\right)_1+\left(\mathcal{J}_{4+}^{(1)}\right)_1 ,
		\end{aligned}
	\end{equation}
	where, from \eqref{E3E4},
	\begin{equation}
		\begin{aligned}
			\left(J_{1+}^{(1)}\right)_1&=m_1\mathcal{A}_3\mathcal{B}_4 \int_{\mathbb{R}^3\times\mathbb{R}^3\times\mathbb{S}^2}H(g^2-\delta_{12}^{34})g\widehat{\sigma}_{1234}(g,\widehat{\Omega}\cdot\widehat{\Omega}^\prime)  \\
			&\hspace{1.5cm}\times\left[\left(\mathbf{G}^*-\alpha_{34}\sqrt{\dfrac{\mu_{12}}{\mu_{34}}(g^2-\delta_{12}^{34})}\widehat{\Omega}^\prime\right)\cdot\mathbf{u}_4^*\right]\left(G^*_1+\alpha_{21}g_1\right)\\
			&\hspace{1.5cm}\times\exp\left[-\dfrac{m_1+m_2}{2T}(G^*)^2-\dfrac{\mu_{12}}{2T}g^2\right]d\mathbf{G}^*d\mathbf{g}d\widehat{\Omega}^\prime\\
			&=m_1\mathcal{A}_3\mathcal{B}_4 \int_{\mathbb{R}^3\times\mathbb{R}^3\times\mathbb{S}^2}H(g^2-\delta_{12}^{34})g\widehat{\sigma}_{1234}(g,\widehat{\Omega}\cdot\widehat{\Omega}^\prime)(G_1^*)^2u_{41}^*  \\
			&\hspace{1.5cm}\times\exp\left[-\dfrac{m_1+m_2}{2T}(G^*)^2-\dfrac{\mu_{12}}{2T}g^2\right]d\mathbf{G}^*d\mathbf{g}d\widehat{\Omega}^\prime\\
			&=m_1\mathcal{A}_3\mathcal{B}_4\widehat{\lambda}_{1234}^{(0)} I_{100}^0Y_{000}^0u_{41}^*
		\end{aligned}
	\end{equation}
	\begin{equation}
		\begin{aligned}
			\left(J_{2+}^{(1)}\right)_1&=m_1\mathcal{A}_4\mathcal{B}_3 \int_{\mathbb{R}^3\times\mathbb{R}^3\times\mathbb{S}^2}H(g^2-\delta_{12}^{34})g\widehat{\sigma}_{1234}(g,\widehat{\Omega}\cdot\widehat{\Omega}^\prime)  \\
			&\hspace{1.5cm}\times\left[\left(\mathbf{G}^*+\alpha_{43}\sqrt{\dfrac{\mu_{12}}{\mu_{34}}(g^2-\delta_{12}^{34})}\widehat{\Omega}^\prime\right)\cdot\mathbf{u}_3^*\right]\left(G^*_1+\alpha_{21}g_1\right)\\
			&\hspace{1.5cm}\times\exp\left[-\dfrac{m_1+m_2}{2T}(G^*)^2-\dfrac{\mu_{12}}{2T}g^2\right]d\mathbf{G}^*d\mathbf{g}d\widehat{\Omega}^\prime\\
			&=m_1\mathcal{A}_4\mathcal{B}_3 \int_{\mathbb{R}^3\times\mathbb{R}^3\times\mathbb{S}^2}H(g^2-\delta_{12}^{34})g\widehat{\sigma}_{1234}(g,\widehat{\Omega}\cdot\widehat{\Omega}^\prime)(G_1^*)^2u_{31}^*  \\
			&\hspace{1.5cm}\times\exp\left[-\dfrac{m_1+m_2}{2T}(G^*)^2-\dfrac{\mu_{12}}{2T}g^2\right]d\mathbf{G}^*d\mathbf{g}d\widehat{\Omega}^\prime\\
			&=m_1\mathcal{A}_4\mathcal{B}_3\widehat{\lambda}_{1234}^{(0)} I_{100}^0Y_{000}^0u_{31}^*
		\end{aligned}
	\end{equation}
	\begin{equation}
		\begin{aligned}
			\left(J_{3+}^{(1)}\right)_1&=m_1\mathcal{B}_3\mathcal{C}_4 \int_{\mathbb{R}^3\times\mathbb{R}^3\times\mathbb{S}^2}H(g^2-\delta_{12}^{34})g\widehat{\sigma}_{1234}(g,\widehat{\Omega}\cdot\widehat{\Omega}^\prime)  \\
			&\hspace{1.5cm}\times\left[\left(\mathbf{G}^*+\alpha_{43}\sqrt{\dfrac{\mu_{12}}{\mu_{34}}(g^2-\delta_{12}^{34})}\widehat{\Omega}^\prime\right)\cdot\mathbf{u}_3^*\right] \\
			&\hspace{1.5cm}\times\left[(G^*)^2-2\alpha_{34}\sqrt{\dfrac{\mu_{12}}{\mu_{34}}(g^2-\delta_{12}^{34})}\mathbf{G}^*\cdot\widehat{\Omega}^\prime +\alpha_{34}^2\dfrac{\mu_{12}}{\mu_{34}}(g^2-\delta_{12}^{34})\right] \\
			&\hspace{1.5cm}\times\left(G^*_1+\alpha_{21}g_1\right)\exp\left[-\dfrac{m_1+m_2}{2T}(G^*)^2-\dfrac{\mu_{12}}{2T}g^2\right]d\mathbf{G}^*d\mathbf{g}d\widehat{\Omega}^\prime\\
			&=m_1\mathcal{B}_3\mathcal{C}_4 \int_{\mathbb{R}^3\times\mathbb{R}^3\times\mathbb{S}^2}H(g^2-\delta_{12}^{34})g\widehat{\sigma}_{1234}(g,\widehat{\Omega}\cdot\widehat{\Omega}^\prime)  \\
			&\hspace{1.5cm}\times\left[(G_1^*)^2(G^*)^2u_{31}^*+\alpha_{34}^2\dfrac{\mu_{12}}{\mu_{34}} (G_1^*)^2(g^2-\delta_{12}^{34})u_{31}^*\right.\\
			&\hspace{1.5cm}\left.-2\alpha_{12}\alpha_{21} (G_1^*)^2(g^2-\delta_{12}^{34})((\widehat{\Omega}^\prime\otimes\widehat{\Omega}^\prime)\cdot\mathbf{u}_{3}^*)_1\right] \\
			&\hspace{1.5cm}\times\exp\left[-\dfrac{m_1+m_2}{2T}(G^*)^2-\dfrac{\mu_{12}}{2T}g^2\right]d\mathbf{G}^*d\mathbf{g}d\widehat{\Omega}^\prime\\
			&=m_1\mathcal{B}_3\mathcal{C}_4\left\{\widehat{\lambda}_{1234}^{(0)}\left[ I_{100}^1Y_{000}^0+\alpha_{34}^2\dfrac{\mu_{12}}{\mu_{34}}I_{100}^0\left(Y_{000}^1-\delta_{12}^{34}Y_{000}^0\right)\right]u_{31}^*\right.\\
			&\hspace{1.5cm}\left.-2\alpha_{12}\alpha_{21}I_{100}^0\left(Y_{000}^1-\delta_{12}^{34}Y_{000}^0\right)(\widehat{\boldsymbol{\Lambda}}_{1234}^{(2)}\cdot\mathbf{u}_{3}^*)_1\right\}
		\end{aligned}
	\end{equation}
	\begin{equation}
		\begin{aligned}
			\left(J_{4+}^{(1)}\right)_1&=m_1\mathcal{B}_4\mathcal{C}_3 \int_{\mathbb{R}^3\times\mathbb{R}^3\times\mathbb{S}^2}H(g^2-\delta_{12}^{34})g\widehat{\sigma}_{1234}(g,\widehat{\Omega}\cdot\widehat{\Omega}^\prime)  \\
			&\hspace{1.5cm}\times\left[\left(\mathbf{G}^*-\alpha_{34}\sqrt{\dfrac{\mu_{12}}{\mu_{34}}(g^2-\delta_{12}^{34})}\widehat{\Omega}^\prime\right)\cdot\mathbf{u}_4^*\right] \\
			&\hspace{1.5cm}\times\left[(G^*)^2+2\alpha_{43}\sqrt{\dfrac{\mu_{12}}{\mu_{34}}(g^2-\delta_{12}^{34})}\mathbf{G}^*\cdot\widehat{\Omega}^\prime +\alpha_{43}^2\dfrac{\mu_{12}}{\mu_{34}}(g^2-\delta_{12}^{34})\right] \\
			&\hspace{1.5cm}\times\left(G^*_1+\alpha_{21}g_1\right)\exp\left[-\dfrac{m_1+m_2}{2T}(G^*)^2-\dfrac{\mu_{12}}{2T}g^2\right]d\mathbf{G}^*d\mathbf{g}d\widehat{\Omega}^\prime\\
			&=m_1\mathcal{B}_4\mathcal{C}_3 \int_{\mathbb{R}^3\times\mathbb{R}^3\times\mathbb{S}^2}H(g^2-\delta_{12}^{34})g\widehat{\sigma}_{1234}(g,\widehat{\Omega}\cdot\widehat{\Omega}^\prime)  \\
			&\hspace{1.5cm}\times\left[(G_1^*)^2(G^*)^2u_{41}^*+\alpha_{43}^2\dfrac{\mu_{12}}{\mu_{34}} (G_1^*)^2(g^2-\delta_{12}^{34})u_{41}^*\right.\\
			&\hspace{1.5cm}\left.-2\alpha_{12}\alpha_{21} (G_1^*)^2(g^2-\delta_{12}^{34})((\widehat{\Omega}^\prime\otimes\widehat{\Omega}^\prime)\cdot\mathbf{u}_{4}^*)_1\right] \\
			&\hspace{1.5cm}\times\exp\left[-\dfrac{m_1+m_2}{2T}(G^*)^2-\dfrac{\mu_{12}}{2T}g^2\right]d\mathbf{G}^*d\mathbf{g}d\widehat{\Omega}^\prime\\
			&=m_1\mathcal{B}_4\mathcal{C}_3\left\{\widehat{\lambda}_{1234}^{(0)}\left[ I_{100}^1Y_{000}^0+\alpha_{34}^2\dfrac{\mu_{12}}{\mu_{34}}I_{100}^0\left(Y_{000}^1-\delta_{12}^{34}Y_{000}^0\right)\right]u_{41}^*\right.\\
			&\hspace{1.5cm}\left.-2\alpha_{12}\alpha_{21}I_{100}^0\left(Y_{000}^1-\delta_{12}^{34}Y_{000}^0\right)(\widehat{\boldsymbol{\Lambda}}_{1234}^{(2)}\cdot\mathbf{u}_{4}^*)_1\right\}
		\end{aligned}
	\end{equation}
	We can proceed analogously for the other components of the vector $\mathcal{R}_{1+}^{(1)}$.\\
	For what concerns the loss term $\mathcal{R}_{1-}^{(1)}$, the first component is given by
	\begin{equation}
		\begin{aligned}
			\left(\mathcal{R}_{1-}^{(1)}\right)_1 &=\left(\mathcal{J}_{1-}^{(1)}\right)_1+\left(\mathcal{J}_{2-}^{(1)}\right)_1+\left(\mathcal{J}_{3-}^{(1)}\right)_1+\left(\mathcal{J}_{4-}^{(1)}\right)_1 ,
		\end{aligned}
	\end{equation}
	where, from \eqref{E1E2},
	\begin{equation}
		\begin{aligned}
			\left(J_{1-}^{(1)}\right)_1&=m_1\mathcal{A}_1\mathcal{B}_2 \int_{\mathbb{R}^3\times\mathbb{R}^3\times\mathbb{S}^2}H(g^2-\delta_{12}^{34})g\widehat{\sigma}_{1234}(g,\widehat{\Omega}\cdot\widehat{\Omega}^\prime)  \\
			&\hspace{1.5cm}\times\left[\left(\mathbf{G}^*-\alpha_{12}\mathbf{g}^\prime\right)\cdot\mathbf{u}_2^*\right]\left(G^*_1+\alpha_{21}g_1\right)\\
			&\hspace{1.5cm}\times\exp\left[-\dfrac{m_1+m_2}{2T}(G^*)^2-\dfrac{\mu_{12}}{2T}g^2\right]d\mathbf{G}^*d\mathbf{g}d\widehat{\Omega}^\prime\\
			&=m_1\mathcal{A}_1\mathcal{B}_2 \! \int_{\mathbb{R}^3\times\mathbb{R}^3\times\mathbb{S}^2} \!\! H(g^2-\delta_{12}^{34})g\widehat{\sigma}_{1234}(g,\widehat{\Omega}\cdot\widehat{\Omega}^\prime)\left[(G_1^*)^2u_{21}^*-\alpha_{12}\alpha_{21}g_1^2u_{21}\right]  \\
			&\hspace{1.5cm}\times\exp\left[-\dfrac{m_1+m_2}{2T}(G^*)^2-\dfrac{\mu_{12}}{2T}g^2\right]d\mathbf{G}^*d\mathbf{g}d\widehat{\Omega}^\prime\\
			&=m_1\mathcal{A}_1\mathcal{B}_2\widehat{\lambda}_{1234}^{(0)}\left( I_{100}^0Y_{000}^0-\alpha_{12}\alpha_{21}I_{000}^0Y_{100}^0\right)u_{41}^*
		\end{aligned}
	\end{equation}
	\begin{equation}
		\begin{aligned}
			\left(J_{2-}^{(1)}\right)_1&=m_1\mathcal{A}_2\mathcal{B}_1 \int_{\mathbb{R}^3\times\mathbb{R}^3\times\mathbb{S}^2}H(g^2-\delta_{12}^{34})g\widehat{\sigma}_{1234}(g,\widehat{\Omega}\cdot\widehat{\Omega}^\prime)  \\
			&\hspace{1.5cm}\times\left[\left(\mathbf{G}^*+ \alpha_{21}\mathbf{g}^\prime\right)\cdot\mathbf{u}_2^*\right]\left(G^*_1+\alpha_{21}g_1\right)\\
			&\hspace{1.5cm}\times\exp\left[-\dfrac{m_1+m_2}{2T}(G^*)^2-\dfrac{\mu_{12}}{2T}g^2\right]d\mathbf{G}^*d\mathbf{g}d\widehat{\Omega}^\prime\\
			&=m_1\mathcal{A}_2\mathcal{B}_1 \int_{\mathbb{R}^3\times\mathbb{R}^3\times\mathbb{S}^2}H(g^2-\delta_{12}^{34})g\widehat{\sigma}_{1234}(g,\widehat{\Omega}\cdot\widehat{\Omega}^\prime)\left[(G_1^*)^2u_{21}^*+ \alpha_{21}^2g_1^2u_{21}\right]  \\
			&\hspace{1.5cm}\times\exp\left[-\dfrac{m_1+m_2}{2T}(G^*)^2-\dfrac{\mu_{12}}{2T}g^2\right]d\mathbf{G}^*d\mathbf{g}d\widehat{\Omega}^\prime\\
			&=m_1\mathcal{A}_2\mathcal{B}_1\widehat{\lambda}_{1234}^{(0)}\left( I_{100}^0Y_{000}^0+\alpha_{21}^2I_{000}^0Y_{100}^0\right)u_{41}^*
		\end{aligned}
	\end{equation}
	\begin{equation}
		\begin{aligned}
			\left(J_{3-}^{(1)}\right)_1&=m_1\mathcal{B}_1\mathcal{C}_2 \int_{\mathbb{R}^3\times\mathbb{R}^3\times\mathbb{S}^2}H(g^2-\delta_{12}^{34})g\widehat{\sigma}_{1234}(g,\widehat{\Omega}\cdot\widehat{\Omega}^\prime)  \\
			&\hspace{1.5cm}\times\left[\left(\mathbf{G}^*+\alpha_{21}\mathbf{g}\right)\cdot\mathbf{u}_1^*\right]\left[(G^*)^2-2\alpha_{12}\mathbf{G}^*\cdot\mathbf{g} +\alpha_{12}^2g^2\right] \\
			&\hspace{1.5cm}\times\left(G^*_1+\alpha_{21}g_1\right)\exp\left[-\dfrac{m_1+m_2}{2T}(G^*)^2-\dfrac{\mu_{12}}{2T}g^2\right]d\mathbf{G}^*d\mathbf{g}d\widehat{\Omega}^\prime\\
			&=m_1\mathcal{B}_1\mathcal{C}_2 \int_{\mathbb{R}^3\times\mathbb{R}^3\times\mathbb{S}^2}H(g^2-\delta_{12}^{34})g\widehat{\sigma}_{1234}(g,\widehat{\Omega}\cdot\widehat{\Omega}^\prime)  \\
			&\hspace{1.5cm}\times\left[(G_1^*)^2(G^*)^2u_{11}^*-4\alpha_{12}\alpha_{21}(G_1^*)^2g_1^2u_{11}^*+\alpha_{12}^2(G_1^*)^2g^2u_{11}^*\right.\\
			&\hspace{1.5cm}\left.+\alpha_{21}^2(G^*)^2g_1^2u_{11}^*+\alpha_{12}^2\alpha_{21}^2g_1^2g^2u_{11}^*\right] \\
			&\hspace{1.5cm}\times\exp\left[-\dfrac{m_1+m_2}{2T}(G^*)^2-\dfrac{\mu_{12}}{2T}g^2\right]d\mathbf{G}^*d\mathbf{g}d\widehat{\Omega}^\prime\\
			&=m_1\mathcal{B}_1\mathcal{C}_2\widehat{\lambda}_{1234}^{(0)}\left(I_{100}^1Y_{000}^0-4\alpha_{12}\alpha_{21}I_{100}^0Y_{100}^0+\alpha_{12}^2I_{100}^0Y_{000}^1\right.\\
			&\hspace{1.5cm}\left.+\alpha_{21}^2I_{000}^1Y_{100}^0+\alpha_{12}^2\alpha_{21}^2I_{000}^0Y_{100}^1\right)u_{11}^*
		\end{aligned}
	\end{equation}
	\begin{equation}
		\begin{aligned}
			\left(J_{4-}^{(1)}\right)_1&=m_1\mathcal{B}_2\mathcal{C}_1 \int_{\mathbb{R}^3\times\mathbb{R}^3\times\mathbb{S}^2}H(g^2-\delta_{12}^{34})g\widehat{\sigma}_{1234}(g,\widehat{\Omega}\cdot\widehat{\Omega}^\prime)  \\
			&\hspace{1.5cm}\times\left[\left(\mathbf{G}^*-\alpha_{12}\mathbf{g}\right)\cdot\mathbf{u}_1^*\right]\left[(G^*)^2+2\alpha_{21}\mathbf{G}^*\cdot\mathbf{g} +\alpha_{21}^2g^2\right] \\
			&\hspace{1.5cm}\times\left(G^*_1+\alpha_{21}g_1\right)\exp\left[-\dfrac{m_1+m_2}{2T}(G^*)^2-\dfrac{\mu_{12}}{2T}g^2\right]d\mathbf{G}^*d\mathbf{g}d\widehat{\Omega}^\prime\\
			&=m_1\mathcal{B}_2\mathcal{C}_1 \int_{\mathbb{R}^3\times\mathbb{R}^3\times\mathbb{S}^2}H(g^2-\delta_{12}^{34})g\widehat{\sigma}_{1234}(g,\widehat{\Omega}\cdot\widehat{\Omega}^\prime)  \\
			&\hspace{1.5cm}\times\left[(G_1^*)^2(G^*)^2u_{11}^* \!+\! 2\alpha_{21}(\alpha_{21}  \!-\! \alpha_{12})(G_1^*)^2g_1^2u_{11}^*
			\!+\! \alpha_{21}^2(G_1^*)^2g^2u_{11}^*\right.\\
			&\hspace{1.5cm}\left.-\alpha_{12}\alpha_{21}(G^*)^2g_1^2u_{11}^*-\alpha_{12}\alpha_{21}^3g_1^2g^2u_{11}^*\right] \\
			&\hspace{1.5cm}\times\exp\left[-\dfrac{m_1+m_2}{2T}(G^*)^2-\dfrac{\mu_{12}}{2T}g^2\right]d\mathbf{G}^*d\mathbf{g}d\widehat{\Omega}^\prime\\
			&=m_1\mathcal{B}_2\mathcal{C}_1\widehat{\lambda}_{1234}^{(0)}\left[I_{100}^1Y_{000}^0+2\alpha_{21}(\alpha_{21}-\alpha_{12})I_{100}^0Y_{100}^0+\alpha_{21}^2I_{100}^0Y_{000}^1\right.\\
			&\hspace{1.5cm}\left.-\alpha_{12}\alpha_{21}I_{000}^1Y_{100}^0+\alpha_{12}\alpha_{21}^3I_{000}^0Y_{100}^1\right]u_{11}^*
		\end{aligned}
	\end{equation}
	Similar expressions can be deduced for the other constituents.
	
	
	\subsection{Production of energy}
	\label{ssec:en}
	
	We use now the test function $\varphi_1^{(2)}=\dfrac12m_1(v^*)^2$; the gain term can be written as
	\begin{equation}
		\begin{aligned}
			\mathcal{R}_{1+}^{(2)} &=\mathcal{J}_{1+}^{(2)}+\mathcal{J}_{2+}^{(2)}+\mathcal{J}_{3+}^{(2)}+\mathcal{J}_{4+}^{(2)}+\mathcal{J}_{5+}^{(2)}\,,
		\end{aligned}
	\end{equation}
	where
	\begin{equation}
		\begin{aligned}
			J_{1+}^{(2)}&=\dfrac12m_1\mathcal{A}_3\mathcal{A}_4 \int_{\mathbb{R}^3\times\mathbb{R}^3\times\mathbb{S}^2}H(g^2-\delta_{12}^{34})g\widehat{\sigma}_{1234}(g,\widehat{\Omega}\cdot\widehat{\Omega}^\prime) \\
			&\hspace{1.0cm}\times\left[(G^*)^2  \!+\! 2\alpha_{21}\mathbf{G}^*\cdot\mathbf{g}  \!+\! \alpha_{21}^2g^2\right]
			\exp \! \left[-\dfrac{m_1  \!+\! m_2}{2T}(G^*)^2  \!-\! \dfrac{\mu_{12}}{2T}g^2\right] d\mathbf{G}^*\! d\mathbf{g}d\widehat{\Omega}^\prime\\
			&=\dfrac12m_1\mathcal{A}_3\mathcal{A}_4\widehat{\lambda}_{1234}^{(0)} \left(I_{000}^1Y_{000}^0+\alpha_{21}^2I_{000}^0Y_{000}^1\right)
		\end{aligned}
	\end{equation}
	\begin{equation}
		\begin{aligned}
			J_{2+}^{(2)}&=\dfrac12m_1\mathcal{A}_3\mathcal{C}_4 \int_{\mathbb{R}^3\times\mathbb{R}^3\times\mathbb{S}^2}H(g^2-\delta_{12}^{34})g\widehat{\sigma}_{1234}(g,\widehat{\Omega}\cdot\widehat{\Omega}^\prime) \\
			&\hspace{1.0cm}\times\left[(G^*)^2-2\alpha_{34}\sqrt{\dfrac{\mu_{12}}{\mu_{34}}(g^2-\delta_{12}^{34})}\mathbf{G}^*\cdot\widehat{\Omega}^\prime+\alpha_{34}^2\dfrac{\mu_{12}}{\mu_{34}}(g^2-\delta_{12}^{34})\right]\\
			&\hspace{1.0cm}\times\left[(G^*)^2  \!+\! 2\alpha_{21}\mathbf{G}^*\cdot\mathbf{g}  \!+\! \alpha_{21}^2g^2\right]
			\exp\!\left[\!-\dfrac{m_1 \!+\!m_2}{2T}(G^*)^2  \!-\! \dfrac{\mu_{12}}{2T}g^2\right] \! d\mathbf{G}^* \! d\mathbf{g}  d\widehat{\Omega}^\prime\\
			&=\dfrac12m_1\mathcal{A}_3\mathcal{C}_4 \int_{\mathbb{R}^3\times\mathbb{R}^3\times\mathbb{S}^2}H(g^2-\delta_{12}^{34})g\widehat{\sigma}_{1234}(g,\widehat{\Omega}\cdot\widehat{\Omega}^\prime) \\
			&\hspace{1.0cm}\times \! \left[(G^*)^4  \!+\! \alpha_{21}^2(G^*)^2g^2  \!+\! \alpha_{34}^2(G^*)^2 \! \dfrac{\mu_{12}}{\mu_{34}}(g^2  \!-\! \delta_{12}^{34}) \!+\!\alpha_{21}^2\alpha_{34}^2 \! \dfrac{\mu_{12}}{\mu_{34}}(g^2  \!-\! \delta_{12}^{34})g^2\right]\\
			&\hspace{1.0cm}\times\exp\left[-\dfrac{m_1+m_2}{2T}(G^*)^2-\dfrac{\mu_{12}}{2T}g^2\right]d\mathbf{G}^*d\mathbf{g}d\widehat{\Omega}^\prime\\
			&=\dfrac12m_1\mathcal{A}_3\mathcal{C}_4\widehat{\lambda}_{1234}^{(0)} \left[I_{000}^2Y_{000}^0+\alpha_{21}^2I_{000}^1Y_{000}^1+\alpha_{34}^2\dfrac{\mu_{12}}{\mu_{34}}I_{000}^1(Y_{000}^1-\delta_{12}^{34}Y_{000}^0)\right.\\
			&\hspace{1.0cm}\left.+\alpha_{21}^2\alpha_{34}^2\dfrac{\mu_{12}}{\mu_{34}}I_{000}^0(Y_{000}^2-\delta_{12}^{34}Y_{000}^1)\right]
		\end{aligned}
	\end{equation}
	\begin{equation}
		\begin{aligned}
			J_{3+}^{(2)}&=\dfrac12m_1\mathcal{A}_4\mathcal{C}_3 \int_{\mathbb{R}^3\times\mathbb{R}^3\times\mathbb{S}^2}H(g^2-\delta_{12}^{34})g\widehat{\sigma}_{1234}(g,\widehat{\Omega}\cdot\widehat{\Omega}^\prime) \\
			&\hspace{1.0cm}\times\left[(G^*)^2+2\alpha_{43}\sqrt{\dfrac{\mu_{12}}{\mu_{34}}(g^2-\delta_{12}^{34})}\mathbf{G}^*\cdot\widehat{\Omega}^\prime+\alpha_{43}^2\dfrac{\mu_{12}}{\mu_{34}}(g^2-\delta_{12}^{34})\right]\\
			&\hspace{1.0cm}\times\left[(G^*)^2  \!+\! 2\alpha_{21}\mathbf{G}^*\cdot\mathbf{g}  \!+\! \alpha_{21}^2g^2\right] \!
			\exp \! \left[ \! -\dfrac{m_1  \!+\! m_2}{2T}(G^*)^2  \!-\! \dfrac{\mu_{12}}{2T}g^2\right] \!
			d\mathbf{G}^*d\mathbf{g}d\widehat{\Omega}^\prime\\
			&=\dfrac12m_1\mathcal{A}_4\mathcal{C}_3 \int_{\mathbb{R}^3\times\mathbb{R}^3\times\mathbb{S}^2}H(g^2-\delta_{12}^{34})g\widehat{\sigma}_{1234}(g,\widehat{\Omega}\cdot\widehat{\Omega}^\prime) \\
			&\hspace{1.0cm}\times\left[(G^*)^4  \!+\! \alpha_{21}^2(G^*)^2g^2  \!+\!\alpha_{43}^2(G^*)^2 \!
			\dfrac{\mu_{12}}{\mu_{34}}(g^2  \!-\! \delta_{12}^{34})  \!+\! \alpha_{21}^2\alpha_{43}^2\dfrac{\mu_{12}}{\mu_{34}}(g^2  \!+\! \delta_{12}^{34})g^2\right]\\
			&\hspace{1.0cm}\times\exp\left[-\dfrac{m_1+m_2}{2T}(G^*)^2-\dfrac{\mu_{12}}{2T}g^2\right]d\mathbf{G}^*d\mathbf{g}d\widehat{\Omega}^\prime\\
			&=\dfrac12m_1\mathcal{A}_4\mathcal{C}_3\widehat{\lambda}_{1234}^{(0)} \left[I_{000}^2Y_{000}^0+\alpha_{21}^2I_{000}^1Y_{000}^1+\alpha_{43}^2\dfrac{\mu_{12}}{\mu_{34}}I_{000}^1(Y_{000}^1-\delta_{12}^{34}Y_{000}^0)\right.\\
			&\hspace{1.0cm}\left.+\alpha_{21}^2\alpha_{43}^2\dfrac{\mu_{12}}{\mu_{34}}I_{000}^0(Y_{000}^2-\delta_{12}^{34}Y_{000}^1)\right]
		\end{aligned}
	\end{equation}
	\begin{equation}
		\begin{aligned}
			J_{4+}^{(2)}&=\dfrac12m_1\mathcal{B}_3\mathcal{B}_4 \int_{\mathbb{R}^3\times\mathbb{R}^3\times\mathbb{S}^2}H(g^2-\delta_{12}^{34})g\widehat{\sigma}_{1234}(g,\widehat{\Omega}\cdot\widehat{\Omega}^\prime) \\
			&\hspace{1.0cm}\times \! \left[ \! (\mathbf{G}^* \!+\! \alpha_{43}\sqrt{\dfrac{\mu_{12}}{\mu_{34}}(g^2  \!-\! \delta_{12}^{34})}\widehat{\Omega}^\prime)\cdot\mathbf{u}_3^* \! \right] \! \left[ \! (\mathbf{G}^*  \!-\! \alpha_{34}\sqrt{\dfrac{\mu_{12}}{\mu_{34}}(g^2  \!-\! \delta_{12}^{34})}
			\widehat{\Omega}^\prime) \cdot \mathbf{u}_4^*\right]\\ 
			&\hspace{1.0cm}\times\left[(G^*)^2  \!+\! 2\alpha_{21}\mathbf{G}^*\cdot\mathbf{g}  \!+\! \alpha_{21}^2g^2\right] \!
			\exp \! \left[ \! -\dfrac{m_1  \!+\! m_2}{2T}(G^*)^2  \!-\! \dfrac{\mu_{12}}{2T}g^2\right] \! d\mathbf{G}^*d\mathbf{g}d\widehat{\Omega}^\prime\\
			&=\dfrac12m_1\mathcal{B}_3\mathcal{B}_4 \int_{\mathbb{R}^3\times\mathbb{R}^3\times\mathbb{S}^2}H(g^2-\delta_{12}^{34})g\widehat{\sigma}_{1234}(g,\widehat{\Omega}\cdot\widehat{\Omega}^\prime) \\
			&\hspace{1.0cm}\left[ \! (G^*)^2 \! \sum_{i=1}^3(G_i^*)^2u_{3i}^*u_{4i}^*  \!+\! \alpha_{21}^2g^2\sum_{i=1}^3(G_i^*)^2u_{3i}^*u_{4i}^*
			\!-\! \alpha_{34}\alpha_{43}(G^*)^2 \! \dfrac{\mu_{12}}{\mu_{34}}(g^2  \!-\! \delta_{12}^{34}) \right.\\
			&\hspace{1.0cm}\left.\times\sum_{i,j=1}^3\widehat{\Omega}_i^\prime\widehat{\Omega}_j^\prime u_{3i}^*u_{4j}^*-\alpha_{34}\alpha_{43}\alpha_{21}^2\dfrac{\mu_{12}}{\mu_{34}}g^2(g^2-\delta_{12}^{34})\sum_{i,j=1}^3\widehat{\Omega}_i^\prime\widehat{\Omega}_j^\prime u_{3i}^*u_{4j}^*\right]\\
			&=\dfrac12m_1\mathcal{B}_3\mathcal{B}_4\left[\widehat{\lambda}_{1234}^{(0)} \left(I_{100}^1u_{31}^*u_{41}^*+I_{010}^1u_{32}^*u_{42}^*+I_{001}^1u_{33}^*u_{43}^*\right)Y_{000}^0\right.\\
			&\hspace{1.0cm}\left.+\widehat{\lambda}_{1234}^{(0)}\alpha_{21}^2 \left(I_{100}^0u_{31}^*u_{41}^*+I_{010}^0u_{32}^*u_{42}^*+I_{001}^0u_{33}^*u_{43}^*\right)Y_{000}^1\right.\\
			&\hspace{1.0cm}\left.-\alpha_{12}\alpha_{21}I_{000}^1(Y_{000}^1-\delta_{12}^{34}Y_{000}^0)\widehat{\boldsymbol{\Lambda}}_{1234}^{(2)}:(\mathbf{u}_3^*\otimes\mathbf{u}_4^*)\right.\\
			&\hspace{1.0cm}\left.-\alpha_{12}\alpha_{21}^3I_{000}^0(Y_{000}^2-\delta_{12}^{34}Y_{000}^1)\widehat{\boldsymbol{\Lambda}}_{1234}^{(2)}:(\mathbf{u}_3^*\otimes\mathbf{u}_4^*)\right]\\
			&=\dfrac12m_1\mathcal{B}_3\mathcal{B}_4\left\{\widehat{\lambda}_{1234}^{(0)}\left(I_{100}^1Y_{000}^0+\alpha_{21}^2I_{100}^0Y_{000}^1\right)\mathbf{u}_3^*\cdot\mathbf{u}_4^*-\right.\\
			&\hspace{1.0cm}\left.
			\alpha_{12}\alpha_{21}\left[I_{000}^1(Y_{000}^1  \!-\! \delta_{12}^{34}Y_{000}^0)  \!+\! \alpha_{21}^2I_{000}^0(Y_{000}^2  \!-\! \delta_{12}^{34}Y_{000}^1)\right]\widehat{\boldsymbol{\Lambda}}_{1234}^{(2)}:(\mathbf{u}_3^*\otimes\mathbf{u}_4^*)\right\}
		\end{aligned}
	\end{equation}
	\begin{equation}
		\begin{aligned}
			J_{5+}^{(2)}&=\dfrac12m_1\mathcal{C}_3\mathcal{C}_4 \int_{\mathbb{R}^3\times\mathbb{R}^3\times\mathbb{S}^2}H(g^2-\delta_{12}^{34})g\widehat{\sigma}_{1234}(g,\widehat{\Omega}\cdot\widehat{\Omega}^\prime) \\
			&\hspace{1.0cm}\times\left[(G^*)^2+2\alpha_{43}\sqrt{\dfrac{\mu_{12}}{\mu_{34}}(g^2-\delta_{12}^{34})}\mathbf{G}^*\cdot\widehat{\Omega}^\prime+\alpha_{43}^2\dfrac{\mu_{12}}{\mu_{34}}(g^2-\delta_{12}^{34})\right] \\
			&\hspace{1.0cm}\times \! \left[ \! (G^*)^2  \!-\! 2\alpha_{34}\sqrt{\dfrac{\mu_{12}}{\mu_{34}}(g^2-\delta_{12}^{34})}\mathbf{G}^*\cdot\widehat{\Omega}^\prime+\alpha_{34}^2\dfrac{\mu_{12}}{\mu_{34}}(g^2-\delta_{12}^{34})\right] \\ &\hspace{1.0cm}\times\left[(G^*)^2+2\alpha_{21}\mathbf{G}^*\cdot\mathbf{g}+\alpha_{21}^2g^2\right]\exp\left[-\dfrac{m_1+m_2}{2T}(G^*)^2-\dfrac{\mu_{12}}{2T}g^2\right]d\mathbf{G}^*d\mathbf{g}d\widehat{\Omega}^\prime\\
			&=\dfrac12m_1\mathcal{C}_3\mathcal{C}_4 \int_{\mathbb{R}^3\times\mathbb{R}^3\times\mathbb{S}^2}H(g^2-\delta_{12}^{34})g\widehat{\sigma}_{1234}(g,\widehat{\Omega}\cdot\widehat{\Omega}^\prime) \\
			&\hspace{1.0cm} \! \times \! \left[ \! (G^*)^6  \!+\! \alpha_{21}^2(G^*)^4g^2  \!+\! \alpha_{34}^2 \! \dfrac{\mu_{12}}{\mu_{34}}(G^*)^4 \!
			(g^2  \!-\! \delta_{12}^{34})  \!+\! \alpha_{21}^2\alpha_{34}^2\dfrac{\mu_{12}}{\mu_{34}}(G^*)^2g^2(g^2  \!-\! \delta_{12}^{34})\right.\\
			&\hspace{1.0cm}\left.-4\alpha_{34}\alpha_{43}\dfrac{\mu_{12}}{\mu_{34}}(G^*)^2(g^2-\delta_{12}^{34})\sum_{i,j=1}^3G^*_iG^*_j\widehat{\Omega}_i^\prime\widehat{\Omega}_j^\prime\right.\\
			&\hspace{1.0cm}\left.-4\alpha_{21}^2\alpha_{34}\alpha_{43}\dfrac{\mu_{12}}{\mu_{34}}g^2(g^2-\delta_{12}^{34})\sum_{i,j=1}^3G^*_iG^*_j\widehat{\Omega}_i^\prime\widehat{\Omega}_j^\prime\right.\\
			&\hspace{1.0cm}\left.+\alpha_{43}^2\dfrac{\mu_{12}}{\mu_{34}}(G^*)^4(g^2-\delta_{12}^{34})+\alpha_{43}^2\dfrac{\mu_{12}}{\mu_{34}}(G^*)^2g^2(g^2-\delta_{12}^{34})\right.\\
			&\hspace{1.0cm}\left.+\alpha_{34}^2\alpha_{43}^2\dfrac{\mu_{12}^2}{\mu_{34}^2}(G^*)^2(g^2-\delta_{12}^{34})^2+\alpha_{21}^2\alpha_{34}^2\alpha_{43}^2\dfrac{\mu_{12}^2}{\mu_{34}^2}g^2(g^2-\delta_{12}^{34})^2\right] \\
			&\hspace{1.0cm}\times \exp\left[-\dfrac{m_1+m_2}{2T}(G^*)^2-\dfrac{\mu_{12}}{2T}g^2\right]d\mathbf{G}^*d\mathbf{g}d\widehat{\Omega}^\prime\\
			&=\dfrac12m_1\mathcal{C}_3\mathcal{C}_4\left[\widehat{\lambda}_{1234}^{(0)}I_{000}^3Y_{000}^0+\widehat{\lambda}_{1234}^{(0)}\alpha_{21}^2I_{000}^2Y_{000}^1+\widehat{\lambda}_{1234}^{(0)}\alpha_{34}^2\dfrac{\mu_{12}}{\mu_{34}}I_{000}^2(Y_{000}^1-\delta_{12}^{34}Y_{000}^0)\right.\\
			&\hspace{1.0cm} \left.  \!+\! \widehat{\lambda}_{1234}^{(0)}\alpha_{21}^2\alpha_{34}^2 \! \dfrac{\mu_{12}}{\mu_{34}}I_{000}^1 \!
			(Y_{000}^2  \!-\! \delta_{12}^{34}Y_{000}^1)  \!-\! 4\alpha_{12}\alpha_{21}I_{100}^1(Y_{000}^1-\delta_{12}^{34}Y_{000}^0)\text{tr}(\widehat{\boldsymbol{\Lambda}}_{1234}^{(0)})\right.\\
			&\hspace{1.0cm}\left.-4\alpha_{12}\alpha_{21}^3I_{100}^0(Y_{000}^2-\delta_{12}^{34}Y_{000}^1)\text{tr}(\widehat{\boldsymbol{\Lambda}}_{1234}^{(0)})+\widehat{\lambda}_{1234}^{(0)}\alpha_{43}^2\dfrac{\mu_{12}}{\mu_{34}}I_{100}^2(Y_{000}^1-\delta_{12}^{34}Y_{000}^0)\right.\\
			&\hspace{1.0cm}\left.  \!+\! \widehat{\lambda}_{1234}^{(0)} \! \alpha_{43}^2 \! \dfrac{\mu_{12}}{\mu_{34}}I_{100}^1 \! (Y_{000}^2  \!-\! \delta_{12}^{34}Y_{000}^1)  \!+\!\widehat{\lambda}_{1234}^{(0)} \! \alpha_{12}^2 \! \alpha_{21}^2I_{100}^1 \! (Y_{000}^2  \!-\! 2\delta_{12}^{34}Y_{000}^1  \!+\! \delta_{12}^{34}Y_{000}^0)\right.\\
			&\hspace{1.0cm}\left. +\widehat{\lambda}_{1234}^{(0)}\alpha_{12}^2\alpha_{21}^4I_{100}^0(Y_{000}^3-2\delta_{12}^{34}Y_{000}^2+\delta_{12}^{34}Y_{000}^1)\right]
		\end{aligned}
	\end{equation}
	Analogously, we can rewrite
	\begin{equation}
		\begin{aligned}
			\mathcal{R}_{1-}^{(2)} &=\mathcal{J}_{1-}^{(2)}+\mathcal{J}_{2-}^{(2)}+\mathcal{J}_{3-}^{(2)}+\mathcal{J}_{4-}^{(2)}+\mathcal{J}_{5-}^{(2)}\,,
		\end{aligned}
	\end{equation}
	where
	\begin{equation}
		\begin{aligned}
			J_{1-}^{(2)}&=\dfrac12m_1\mathcal{A}_1\mathcal{A}_2 \int_{\mathbb{R}^3\times\mathbb{R}^3\times\mathbb{S}^2}H(g^2-\delta_{12}^{34})g\widehat{\sigma}_{1234}(g,\widehat{\Omega}\cdot\widehat{\Omega}^\prime) \\
			&\hspace{1.0cm}\times \! \left[ \! (G^*)^2  \!+\! 2\alpha_{21}\mathbf{G}^*\cdot\mathbf{g}  \!+\! \alpha_{21}^2g^2 \! \right] \! 
			\exp \! \left[ \! -\dfrac{m_1  \!+\! m_2}{2T}(G^*)^2-\dfrac{\mu_{12}}{2T}g^2\right]d\mathbf{G}^*d\mathbf{g}d\widehat{\Omega}^\prime\\
			&=\dfrac12m_1\mathcal{A}_1\mathcal{A}_2\widehat{\lambda}_{1234}^{(0)} \left(I_{000}^1Y_{000}^0+\alpha_{21}^2I_{000}^0Y_{000}^1\right)
		\end{aligned}
	\end{equation}
	\begin{equation}
		\begin{aligned}
			J_{2-}^{(2)}&=\dfrac12m_1\mathcal{A}_1\mathcal{C}_2 \int_{\mathbb{R}^3\times\mathbb{R}^3\times\mathbb{S}^2}H(g^2-\delta_{12}^{34})g\widehat{\sigma}_{1234}(g,\widehat{\Omega}\cdot\widehat{\Omega}^\prime) \\
			&\hspace{1.0cm}\times\left[(G^*)^2-2\alpha_{12}\mathbf{G}^*\cdot\mathbf{g}+\alpha_{12}^2g^2\right]\left[(G^*)^2+2\alpha_{21}\mathbf{G}^*\cdot\mathbf{g}+\alpha_{21}^2g^2\right] \\
			&\hspace{1.0cm}\times\exp\left[-\dfrac{m_1+m_2}{2T}(G^*)^2-\dfrac{\mu_{12}}{2T}g^2\right]d\mathbf{G}^*d\mathbf{g}d\widehat{\Omega}^\prime\\
			&=\dfrac12m_1\mathcal{A}_1\mathcal{C}_2 \int_{\mathbb{R}^3\times\mathbb{R}^3\times\mathbb{S}^2}H(g^2-\delta_{12}^{34})g\widehat{\sigma}_{1234}(g,\widehat{\Omega}\cdot\widehat{\Omega}^\prime) \\
			&\hspace{1.0cm}\times \! \left[ \! (G^*)^4  \!+\! \alpha_{21}^2(G^*)^2g^2  \!-\! 4\alpha_{12}\alpha_{21}\sum_{i=1}^3(G_i^*)^2g_i^2  \!+\! \alpha_{12}^2(G^*)^2g^2
			\!+\! \alpha_{12}^2\alpha_{21}^2g^4\right] \\
			&\hspace{1.0cm}\times\exp\left[-\dfrac{m_1+m_2}{2T}(G^*)^2-\dfrac{\mu_{12}}{2T}g^2\right]d\mathbf{G}^*d\mathbf{g}d\widehat{\Omega}^\prime\\
			&=\dfrac12m_1\mathcal{A}_1\mathcal{C}_2\widehat{\lambda}_{1234}^{(0)} \left[I_{000}^2Y_{000}^0+\alpha_{21}^2I_{000}^1Y_{000}^1-12\alpha_{12}\alpha_{21}I_{100}^0Y_{100}^0+\alpha_{12}^2I_{000}^1Y_{000}^1\right.\\
			&\hspace{1.0cm}\left.+\alpha_{12}^2\alpha_{21}^2I_{000}^0Y_{000}^2\right]
		\end{aligned}
	\end{equation}
	\begin{equation}
		\begin{aligned}
			J_{3-}^{(2)}&=\dfrac12m_1\mathcal{A}_2\mathcal{C}_1 \int_{\mathbb{R}^3\times\mathbb{R}^3\times\mathbb{S}^2}H(g^2-\delta_{12}^{34})g\widehat{\sigma}_{1234}(g,\widehat{\Omega}\cdot\widehat{\Omega}^\prime) \\
			&\hspace{1.0cm}\times \! \left[ \! (G^*)^2  \!+\! 2\alpha_{21}\mathbf{G}^*\cdot\mathbf{g}  \!+\! \alpha_{21}^2g^2\right]^2 
			\exp \! \left[ \! -\dfrac{m_1  \!+\! m_2}{2T}(G^*)^2  \!-\! \dfrac{\mu_{12}}{2T}g^2 \! \right] d\mathbf{G}^*d\mathbf{g}d\widehat{\Omega}^\prime\\
			&=\dfrac12m_1\mathcal{A}_2\mathcal{C}_1 \int_{\mathbb{R}^3\times\mathbb{R}^3\times\mathbb{S}^2}H(g^2-\delta_{12}^{34})g\widehat{\sigma}_{1234}(g,\widehat{\Omega}\cdot\widehat{\Omega}^\prime) \\
			&\hspace{1.0cm}\times\left[(G^*)^4+\alpha_{21}^2(G^*)^2g^2+4\alpha_{21}^2 \sum_{i=1}^3(G_i^*)^2g_i^2+\alpha_{21}^2(G^*)^2g^2+\alpha_{21}^4g^4\right] \\
			&\hspace{1.0cm}\times\exp\left[-\dfrac{m_1+m_2}{2T}(G^*)^2-\dfrac{\mu_{12}}{2T}g^2\right]d\mathbf{G}^*d\mathbf{g}d\widehat{\Omega}^\prime\\
			&=\dfrac12m_1\mathcal{A}_2\mathcal{C}_1\widehat{\lambda}_{1234}^{(0)} \left[I_{000}^2Y_{000}^0+2\alpha_{21}^2I_{000}^1Y_{000}^1+12\alpha_{21}^2I_{100}^0Y_{100}^0+\alpha_{21}^4I_{000}^0Y_{000}^2\right]
		\end{aligned}
	\end{equation}
	\begin{equation}
		\begin{aligned}
			J_{4-}^{(2)}&=\dfrac12m_1\mathcal{B}_1\mathcal{B}_2 \int_{\mathbb{R}^3\times\mathbb{R}^3\times\mathbb{S}^2}H(g^2-\delta_{12}^{34})g\widehat{\sigma}_{1234}(g,\widehat{\Omega}\cdot\widehat{\Omega}^\prime) \\
			&\hspace{1.5cm}\times\left[(\mathbf{G}^*+\alpha_{21}\mathbf{g})\cdot\mathbf{u}_1^*\right]\left[(\mathbf{G}^*-\alpha_{12}\mathbf{g})\cdot\mathbf{u}_2^*\right]\\ &\hspace{1.5cm}\times\left[(G^*)^2+2\alpha_{21}\mathbf{G}^*\cdot\mathbf{g}+\alpha_{21}^2g^2\right]\exp\left[-\dfrac{m_1+m_2}{2T}(G^*)^2-\dfrac{\mu_{12}}{2T}g^2\right]d\mathbf{G}^*d\mathbf{g}d\widehat{\Omega}^\prime\\
			&=\dfrac12m_1\mathcal{B}_1\mathcal{B}_2 \int_{\mathbb{R}^3\times\mathbb{R}^3\times\mathbb{S}^2}H(g^2-\delta_{12}^{34})g\widehat{\sigma}_{1234}(g,\widehat{\Omega}\cdot\widehat{\Omega}^\prime) \\
			&\hspace{1.5cm}\left[(G^*)^2\sum_{i=1}^3(G_i^*)^2u_{3i}^*u_{4i}^*+\alpha_{21}^2g^2\sum_{i=1}^3(G_i^*)^2u_{3i}^*u_{4i}^*-2\alpha_{12}\alpha_{21}\sum_{i=1}^3(G_i^*)^2g_i^2u_{1i}^*u_{2i}^*\right.\\
			&\hspace{1.5cm}\left.+2\alpha_{21}^2\sum_{i=1}^3 (G_i^*)^2g_i^2u_{1i}^*u_{2i}^*-\alpha_{12}\alpha_{21}(G^*)^2\sum_{i=1}^3 g_i^2u_{1i}^*u_{2i}^*-\alpha_{12}\alpha_{21}^3g^2\sum_{i=1}^3 g_i^2u_{1i}^*u_{2i}^*\right]\\
			&=\dfrac12m_1\mathcal{B}_1\mathcal{B}_2\widehat{\lambda}_{1234}^{(0)}\left\{I_{100}^1Y_{000}^0+\alpha_{21}^2I_{100}^0Y_{000}^1+2\alpha_{21}(\alpha_{21}-\alpha_{12})I_{100}^0Y_
			{100}^0\right.\\
			&\hspace{1.5cm}\left.-\alpha_{12}\alpha_{21}I_{000}^1Y_{100}^0-\alpha_{12}\alpha_{21}^3I_{000}^0Y_{100}^1\right\}\mathbf{u}_1^*\cdot\mathbf{u}_2^*
		\end{aligned}
	\end{equation}
	\begin{equation}
		\begin{aligned}
			J_{5-}^{(2)}&=\dfrac12m_1\mathcal{C}_1\mathcal{C}_2 \int_{\mathbb{R}^3\times\mathbb{R}^3\times\mathbb{S}^2}H(g^2-\delta_{12}^{34})g\widehat{\sigma}_{1234}(g,\widehat{\Omega}\cdot\widehat{\Omega}^\prime) \\
			&\hspace{1.5cm}\times\left[(G^*)^2-2\alpha_{12}\mathbf{G}^*\cdot\mathbf{g}+\alpha_{12}^2g^2\right]\left[(G^*)^2+2\alpha_{21}\mathbf{G}^*\cdot\mathbf{g}+\alpha_{21}^2g^2\right]^2 \\ &\hspace{1.5cm}\times\exp\left[-\dfrac{m_1+m_2}{2T}(G^*)^2-\dfrac{\mu_{12}}{2T}g^2\right]d\mathbf{G}^*d\mathbf{g}d\widehat{\Omega}^\prime\\
			&=\dfrac12m_1\mathcal{C}_1\mathcal{C}_2 \int_{\mathbb{R}^3\times\mathbb{R}^3\times\mathbb{S}^2}H(g^2-\delta_{12}^{34})g\widehat{\sigma}_{1234}(g,\widehat{\Omega}\cdot\widehat{\Omega}^\prime) \\
			&\hspace{1.5cm}\times\left[(G^*)^6+(\alpha_{12}^2+2\alpha_{21}^2)(G^*)^4g^2+\alpha_{21}^2(2\alpha_{12}^2+\alpha_{21}^2)(G^*)^2g^4+\alpha_{12}^2\alpha_{21}^2g^6\right.\\
			&\hspace{1.5cm}\left.+4\alpha_{21}(\alpha_{21}-2\alpha_{12})(G^*)^2\sum_{i=1}^3(G_i^*)^2g_i^2+4\alpha_{12}\alpha_{21}^2(\alpha_{12}-2\alpha_{21})g^2\sum_{i=1}^3(G_i^*)^2g_i^2\right] \\
			&\hspace{1.5cm}\times \exp\left[-\dfrac{m_1+m_2}{2T}(G^*)^2-\dfrac{\mu_{12}}{2T}g^2\right]d\mathbf{G}^*d\mathbf{g}d\widehat{\Omega}^\prime\\
			&=\dfrac12m_1\mathcal{C}_1\mathcal{C}_2\widehat{\lambda}_{1234}^{(0)}\left[I_{000}^3Y_{000}^0+(\alpha_{12}^2+2\alpha_{21}^2)I_{000}^2Y_{000}^1+\alpha_{21}^2(2\alpha_{12}^2+\alpha_{21}^2)I_{000}^1Y_{000}^2\right.\\
			&\hspace{1.5cm}\left.+\alpha_{12}^2\alpha_{21}^2I_{000}^0Y_{000}^3 +12\alpha_{21}(\alpha_{21}-2\alpha_{12})I_{100}^1Y_{100}^0+12\alpha_{12}\alpha_{21}^2(\alpha_{12}-2\alpha_{21})I_{100}^0Y_{100}^1\right]
		\end{aligned}
	\end{equation}
	
	
	\section*{Declarations}
	\smallskip
	\noindent
	{\bf Competing interests:} The authors declare no competing interests directly or
	indirectly related to the work submitted for publication.\\
	\noindent
	{\bf Conflict-of-interest statement:} The authors have no conflicts of interest to
	declare.\\
	\noindent
	{\bf Data availability statement:} The authors declare that no datasets were generated or analyzed during the current study.

\end{document}